\documentclass[12pt,preprint]{aastex}

\usepackage{graphicx}
\usepackage{epstopdf}
\usepackage{color}

\slugcomment {light history and evolutionary state of GR~290}

\shorttitle  {GR~290, long term study , evolutionary phase} 
\shortauthors{V. F. Polcaro, O. Maryeva, R. Nesci et al.} 

\begin{document}


\title{GR~290 (Romano's Star): 2. Light history and evolutionary state }


\author{V. F. Polcaro \altaffilmark{1},
O. Maryeva \altaffilmark{2},
R. Nesci \altaffilmark{1},
M. Calabresi\altaffilmark{3},
A. Chieffi\altaffilmark{1},
S. Galleti\altaffilmark{4},
R. Gualandi\altaffilmark{4},
R. Haver\altaffilmark{3},
O. F. Mills\altaffilmark{5},
W. H. Osborn\altaffilmark{5},
A. Pasquali\altaffilmark{6},
C. Rossi\altaffilmark{7},
T. Vasilyeva\altaffilmark{8},
R. F. Viotti\altaffilmark{1} }
\affil{$^1$~INAF-IAPS, Via del Fosso del Cavaliere, 100, 00133 Roma, Italy }
\affil{$^2$~Special Astrophysical Observatory of the Russian Academy of Science, Nizhnii Arkhyz, 369167, Russia}
\affil{$^3$~ARA, Via Carlo Emanuele I, 12A, 00185 Roma, Italy }
\affil{$^4$~INAF - Osservatorio Astronomico di Bologna, Via Ranzani 1, 40127 Bologna, Italy}
\affil{$^5$~ Yerkes Observatory, 373 W. Geneva Street, Williams Bay, WI 53115, USA   }
\affil{ $^6$~Astronomisches Rechen-Institut, Zentrum f\"ur Astronomie, Universit\"at
 Heidelberg, M\"onchhofstrasse 12 - 14, D-69120 Heidelberg, Germany }
\affil{$^7$~Universit\`a La Sapienza, Pza A.Moro 5, 00185 Roma, Italy }
\affil{$^8$~Pulkovo Astronomical Observatory,  196140, Saint-Petersburg, Pulkovskoye chaussee 65/1, Russia }





\begin{abstract}
We have investigated the past light history of the luminous variable~star GR~290 (M33/V532, Romano's Star) in the M~33 galaxy,  
and collected new spectrophotometric observations  
in order to analyse links between this object, the LBV category and the Wolf-Rayet stars of the nitrogen sequence. 
We have built the historical light curve of GR~290 back to 1901, from old observations of the star found in several archival plates of M~33. 
These old recordings together with published and new data on the star 
allowed us to infer that for at least half a century the star was in a low luminosity state, with $B$ $\simeq$18-19, most probably without brighter luminosity phases. 
After 1960, five large variability cycles of visual  luminosity were  recorded.
The amplitude of the oscillations was seen 
increasing towards the 1992-1994 maximum, then decreasing during the last maxima. 
The recent light curve indicates that the photometric variations have been quite similar in all the bands, and that the $B-V$ color index has been constant within $\pm$0.1$^m$ despite the 1.5$^m$ change of the visual luminosity. 
The spectrum of GR~290 at the large maximum of 1992-94, was equivalent to late-B type, while, during 2002-2014, it has varied between WN10h-11h near the  visual  maxima to WN8h-9h at the luminosity minima. We have detected, during this same period, a clear anti-correlation between the visual luminosity, the strength of the \ion{He}{2} 4686~\AA~ emission line, the strength of the 4600-4700~\AA~ lines blend
and the spectral type.
From a model analysis 
of the spectra collected during the whole 2002-2014 period we find  that the Rosseland radius $R_{2/3}$, changed between the minimum and maximum luminosity phases by a factor of 3, while $T_{eff}$ varied between about 33,000~K and 23,000~K. 
We confirm that the bolometric luminosity of the star has not been constant, but has increased by a factor of $\sim$1.5 between minimum and maximum luminosity, in phase with the apparent luminosity variations.
Presently, GR~290 falls in the H-R diagram close to WN8h stars, being probably younger than them. 
In the light of current evolutionary models of very massive stars, we find that GR~290 has evolved from a $\sim$60~$M${\scriptsize $\sun$} progenitor star and should have an age of about 4 million years. From its physical charcteristics, we argue that GR~290 has left the LBV stage and presently 
is moving from the LBV stage to a Wolf-Rayet stage of a late nitrogen spectral type.

\end{abstract} 





\section{Introduction} \label{intro}

GR~290, also known as Romano's star or M33/V532,  in M~33 galaxy was discovered as a variable star in 1978 by \citet{rom78a}, who later described its ample variation as similar to that of Hubble-Sandage variables \citep{rom78b}. 

Extensive studies of GR 290 were performed in the last years, see for example \citet{polc11},  hereafter Paper 1,   \citet{sholu11},  \citet{clark12} and references therein.  
In Paper 1 we have investigated the spectral and photometric variations of GR 290 during the 2003--2010 period and have underlined many important features of the star, namely the spectral type being constantly hotter than for other LBV, and the anti-correlation between the visual luminosity and the excitation of the emission lines. 
In spite of the many studies on GR 290, several questions remain unresolved, for example: the 
light curve so far measured seems peculiar with respect to the well photometrically studied LBVs. 
Furthermore, the hot spectrum of GR 290 is unusual for an LBV and  
be an indication of the star leaving (or having already left) the LBV stage (cf. Paper 1, \citet{hum14} ). 
Recently, a new light minimum, still ongoing
 at the time of writing, has been reported by \citet{cal14}. This has stimulated us to investigate the nature of this star
in more depth.

The aim of this paper is to use the light and spectral variations to trace the past evolution of GR 290, define its
present stage and possible predict its future behavior. 
Following the introduction, the second section will tackle the historical and recent light curve. The third section will be devoted to the spectral variations of the star, while the fourth section will discuss with its evolutionary phase.


\section{The light curve } \label{lightc}
The earliest, targetted observations of GR 290 as performed by \citet{rom78b} covered the period 1960-1977. New observations for the period 1982-1999 were published by \citet{kurtev01}, 
while a very extended list of observations collected at the Crimea Observatory (since 1961) and the Special Astrophysical Observatory (SAO, since 2001) were reported by  \citet{sholu11}. 
These observations, together with  those published in Paper 1 allow us to analyse only the recent "active phase" of the star. 
Since nothing is known about the behavior of GR 290 before 1960, we have searched for available~old plates of the M~33 galaxy 
containing our target.  

\subsection{ Historical plates} \label{histpl}

We have found many  digitised images of  historical plates  obtained by several observatories before 1940. 
 Nine useful plates of M~33  can be found in the Hamburg Observatory archive, taken with the 1 m F/3 Newtonian telescope  The digitized images are available on-line in fits format \footnote{http://plate-archive.hs.uni-hamburg.de/index.php/en/}.
 We  have found also a good number of useful plates taken with the 72 cm F/3.9 reflector between 1907 and 1922
 in the Heidelberg on-line archive \footnote{http://dc.zah.uni-heidelberg.de/lswscans/res/positions/fullplates/form}.
Other plates, found in the Pulkovo Photographic Plate Database \footnote{http://www.puldb.ru/db/plates/index.php} 
\citep{kise07a} were obtained between 1935 and 1938 with the  the 33cm F/11 Normal astrograph of the observatory.
\\ 
Several plates have also been found in the Yerkes Observatory plate archive, most of them obtained with the
Yerkes old 24-inch (60-cm) F/3.87 reflector, exposed between 1901 and 1952. 
Further historic photometric points could be obtained from the POSS plates available~on-line, and from some Asiago 
plates not used by G. Romano in his original paper.

The blue plate of the POSS-I survey is not available on line, thus we only measured the red plate: GR~290 in this plate  is rather faint, outside the range of our photometric sequence, so its magnitude is extrapolated and must be considered rather uncertain. In this case, we included in the comparison sequence the star M31a-314 (R=17.83) to extend the calibration range.  
In all Hamburg plates   GR~290 is clearly visible  except in S02396, although it is generally close to the plate magnitude limit. The bright flux measured in the S02436 plate is uncertain due to the strong coma and the noisy appearance of the emulsion: it would indicate a variation of half magnitude in three months, which however is not unusual for this source as also shown below. 
In several Heidelberg  and Yerkes plates the star is not visible, hence only upper limits  could be derived. In all the plates the star was substantially fainter than the values reported by \citet{rom78b} for the years 1960-62, but is consistent with the magnitude measured during the recent minima. 

All the technical details concerning the data reduction of  the archival images are reported in Appendix A.
Table~\ref{tbl_archmag} lists all the archival magnitudes as derived from our photometric procedure. 

All the old photometric observations are reported in the upper panel of  Figure~\ref{fig_lc}, where we also give the more recent B-filter observations. Some upper limits, that better define the star light curve, have also been shown.

\subsection{New photometry} \label{newobs} 
Since December 2010 new photometric observations
in the B, V and R bands
were collected in Italy with the 1.52 m Cassini telescope at the Loiano Station of the Bologna Astronomical Observatory-INAF, with the 37 cm telescope of Frasso Sabino of the Associazione Romana Astrofili (ARA), and with the 30 cm telescope of Franco Montagni at Greve. We also observed GR~290 with the 80 cm Tenerife telescope of the  Instituto Astrofisico de Canarias (IAC) in October 2012 and October 2013. We made also use of data collected with the 61 cm robotic telescope of the Sierra Stars Observatory Network in Mount Lemmon (Arizona). Table~\ref{tbl_newmag}
reports our new photometric observations, including two U band  values recorded at the Loiano Observatory on November 14 and December 11, 2015.
These new B, V, R observations collected since 2010 are shown, together with the old ones, in the lower panel of Figure~\ref{fig_lc}. 


\subsection{Discussion of the light curve of GR 290} \label{discusLC} 

\begin{figure}
\epsscale{.90}
\plotone{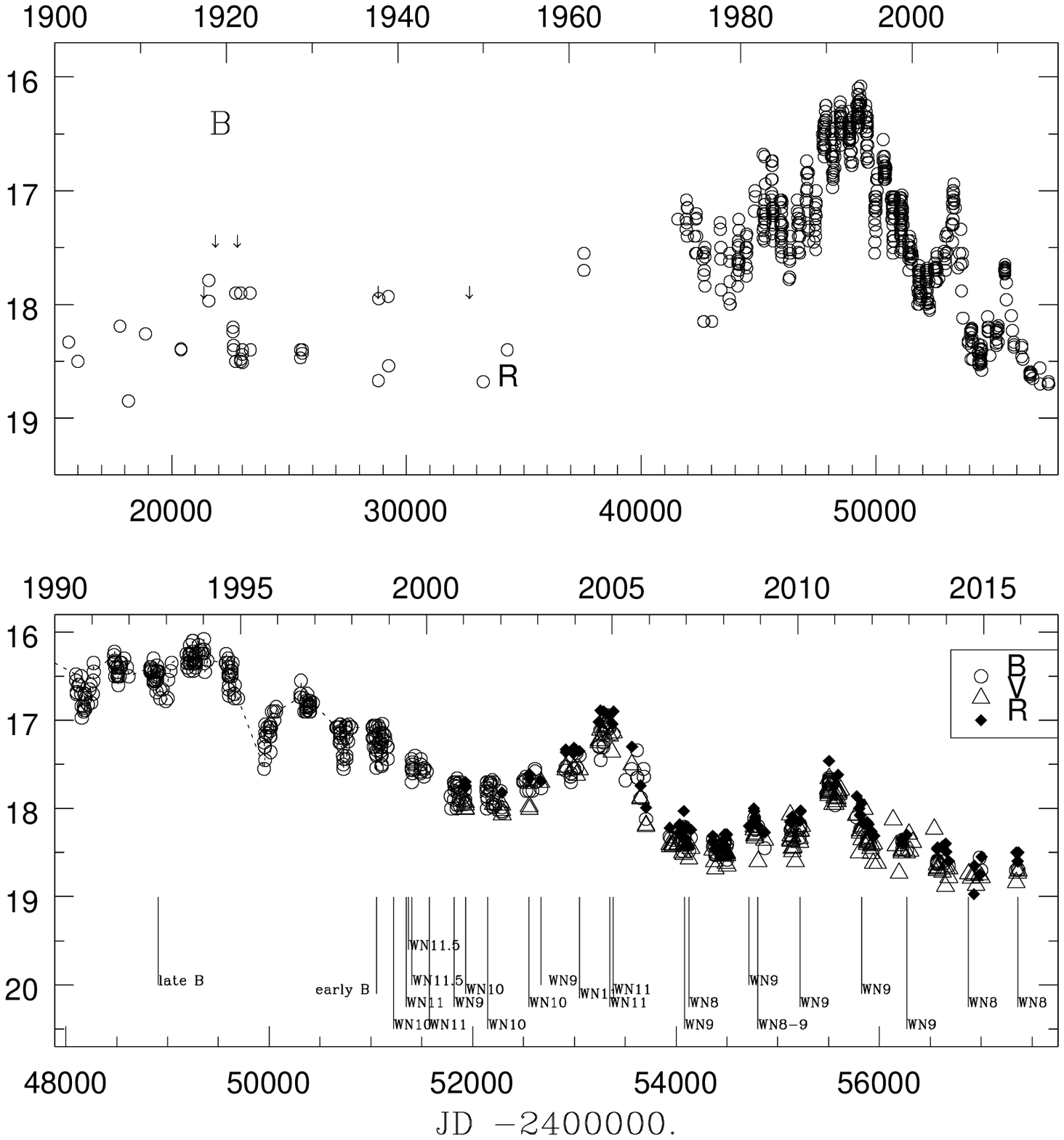}
\caption{ 
Upper panel: The historical light curve of GR 290 in the B-filter from 1901 to 2015. Some upper limits and the R magnitude  from the 1949 POSS plate are also reported.
 Lower panel: a close up of the light curve since the 1990 brightest phase obtained from the B, V, R observations given in Table 3. The spectral types derived from the spectroscopic observations are marked.
 }
 \label{fig_lc}
\end{figure}

As  can be seen in Figure~\ref{fig_lc}, the historical light curve of GR 290 covers a period of more than one century, including a long period of low apparent brightness at 17.5 $<$ B $<$ 18.5 ($\pm$0.2 mag) during the first half of the 20th Century ($quiescent~phase$), followed by a sequence of transitions between low and high  visual luminosity ($active~phase$).
 According to Figure~\ref{fig_lc} one can identify at least five broad light maxima (before 1974, 1983, 1992-94, 2004 and 2010) occurred since 1960, and five light minima (in 1975-79, 1983-86, 2000-2002, 2007-2009 and 2013 to present). 
In particular, the recorded maxima are characterised by different peak luminosities: the first three maxima increased
their peak luminosity, which culminated with the 1992-94 maximum, while the last two occurred with a decreasing peak
luminosity. Also the last three light minima were of
decreasing luminosity. Hence, the photometric trend following the largest 1992-94 maximum
might suggest that the star is gradually recovering a probably new stationary state through
a sequence of oscillations with decreasing  apparent brightness.  As discussed below, this trend is also
marked by the spectroscopic variations of the star. 
In addition, the light curve shows irregular photometric variations on time scales of weeks to months. 
For instance, one can see in  the lower panel of Figure~\ref{fig_lc} that between October 1994 and December 1995 GR~290 was subject to a short event during which its luminosity dropped by at least one magnitude with fairly rapid luminosity decrease and increase. 
No light oscillation of such amplitude has been detected in the recent light curve.

\begin{figure}
\epsscale{.90}
\plotone{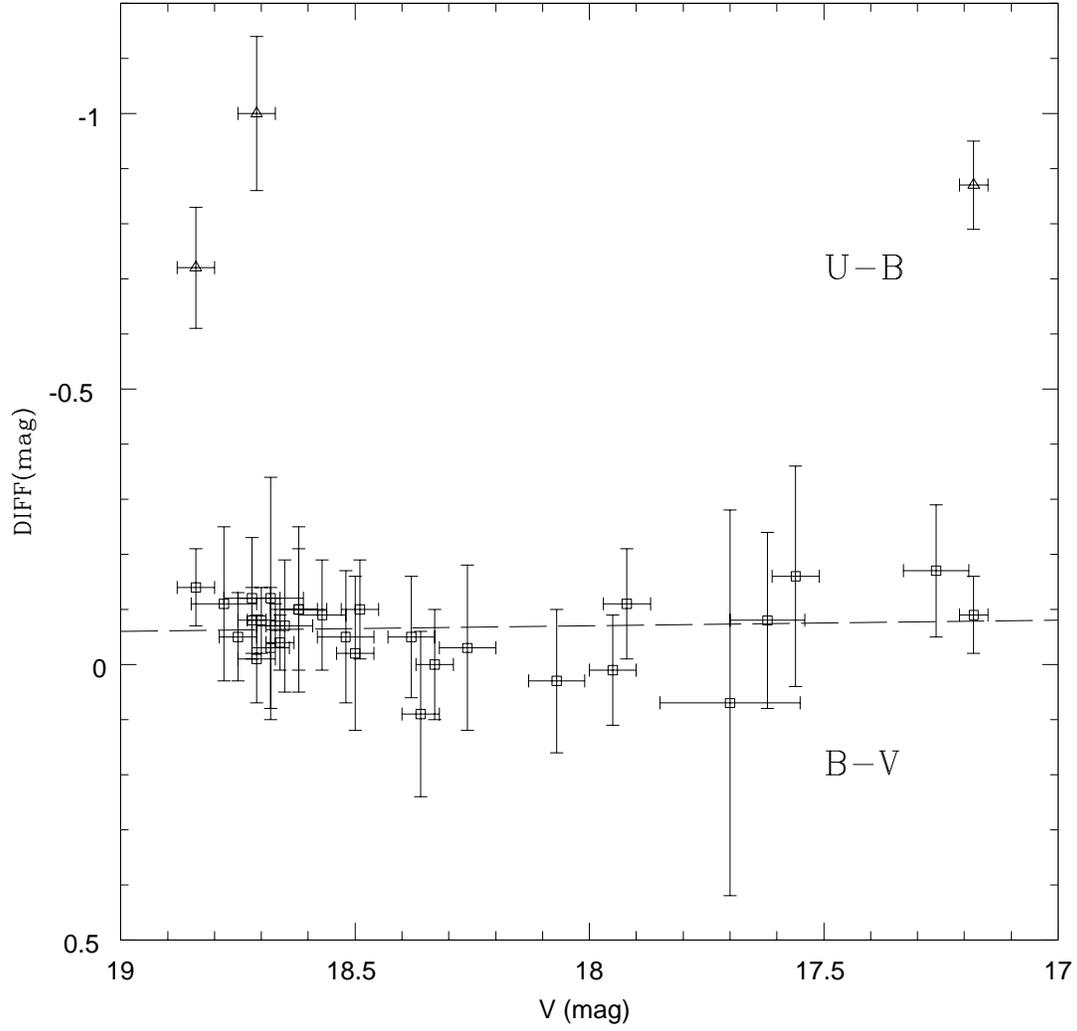}
\caption{ 
 The $(B-V)$ and $(U-B)$ color indices versus the V magnitude
during February 2003-December 2015. The mean $(B-V)$ color index is indicated by the
dotted line.  }
\label{fig_B-VV}
\end{figure}

It is generally observed that, during the S\,Dor phase, the color of LBVs is bluer at the light minimum  
than near the light maximum. In Paper 1 we show that the light curves in the B, V and R bands nearly overlap without a clear luminosity variation (see also Figure 1 in Paper 1). Having now at our disposal a conspicuous amount of multicolor photometric observations covering an ample time interval, we have plotted in Figure~\ref{fig_B-VV} the $(B-V)$ color index versus visual magnitude between February 2003 and December 2014. 

The variations in $(B-V)$ are well within the error bars, with a mean value of about -0.06 mag, and there is no clear evidence for a variation of $(B-V)$ as a function of the visual magnitude. The different behavior of the $(B-V)$ color with the stellar magnitude  given by \citet{sholu11} has to be attributed to the larger uncertainty of their photometry.  
 We conclude that, within the errors,  the $(B-V)$ color index of GR~290 has been basically constant during the last cycles, suggesting that the blue-visual slope of the spectrum has remained substantially unchanged in spite of the ample spectral and  brightness variation.
 This behavior, different from what is generally expected for the S\,Dor variations of LBVs, can be attributed to the continuum opacity which, at the high temperatures of the star, is all the time dominated by electron scattering.
  In fact, in agreement with the observed and theoretical continuum energy distribution of WN stars \citep{morris93}, we also expect  the slope of the blue-visual continuum of GR~290 to be nearly independent of T$_{\rm eff}$ during its recent WN-phase. 
Through the end of 2015 we measured two additional U-band magnitudes: U=17.98 $\pm$ 0.08 (November 14) and  17.78 $\pm$0.1 (December 11). The corresponding $(U-B)$ color index is the same within the errors as that we obtained in December 2004  
 \citep{vio06} when we derived $U$=16.22 and $B$=17.09. The fact that in spite of the large luminosity difference between the two epochs, the blue-ultraviolet color index did not significantly change further supports to the above conclusion.

\section{The spectrum of GR 290} \label{spec} 

The first recorded spectrum of GR~290 was taken by Thomas Szeifert at Calar Alto on October 15, 1992, when the star was near its absolute luminosity maximum with $B$=16.5. 
According to the analysis of Szeifert (1996), this early spectrum is most likely to be of late-B spectral type. However, the next spectrum (1994), obtained in the 5500-7600 \AA\ range at the SAO Russian 6 m telescope, when the star was still in its photometric maximum, led Sholukhova et al. (1997) to suggest that Romano's star could have been a WN, because of the very broad He I 6678 A line.  After the 1992-1994 maximum, GR~290 started to become fainter and hotter.
In September 1998, when the magnitude had already faded to about $B$=17.5, 
the spectrum of GR~290 displayed weak [\ion{N}{2}] emissions and strong \ion{He}{1} lines with P Cygni profiles consistent
with a WN10-WN11 spectral type \citep{fabrika05}.

 The following spectral evolution until 2010 was described in details by    \citet{vio07}, \citet{polc11}, \citet{sholu11}, \citet{mary10} and \citet{hum14}. 
As illustrated in the lower panel of Figure~\ref{fig_lc}, the spectrum of GR 290 was WN9h-WN10h during the 2000-2003 minimum, then became WN11h at the following 2004 maximum. A new hot (WN8h-WN9h) phase was recorded at the 2007 minimum. The subsequent  apparent luminosity increase in 2010 was marked only by a slightly cooler spectrum (WN9h-WN10h), while a new hot (WN8h-WN9h) phase took place during the last minimum phase that is still on-going.
 
New  observations were obtained since 2010 with the BFOSC camera mounted on the 1.52 m Loiano telescope 
and with the SCORPIO camera available with the SAO Russian  6-m telescope \footnote{We used both public data from the SAO RAS archive (http://www.sao.ru/oasis/cgi-bin/fetch) and the data acquired by O. Maryeva during dedicated observations.}.

\begin{figure}
\plotone{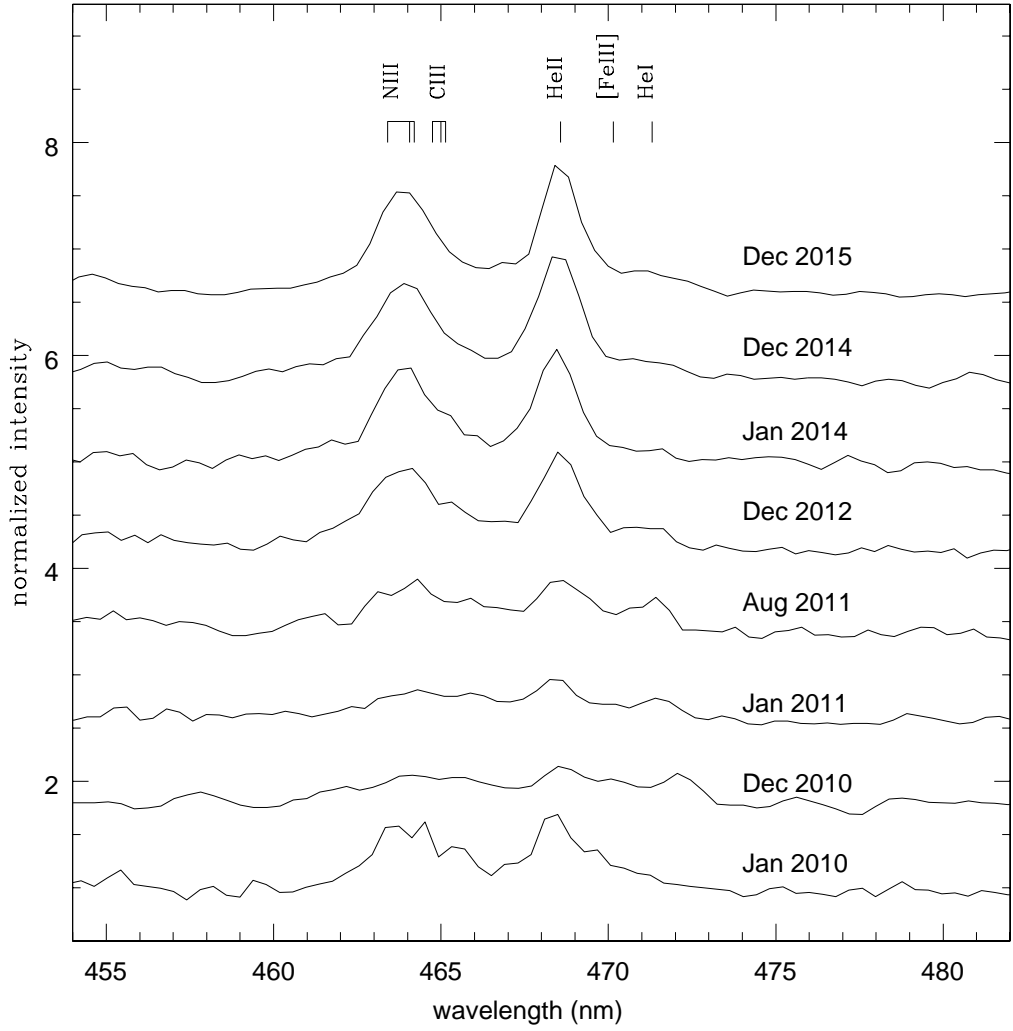}
\caption{ 
The time evolution of the spectrum of GR 290 near the $f$-band between January 2010 and 
December 2015. The successive spectra are vertically shifted.
 }
\label{fig_sp1014}
\end{figure}

In Figure~\ref{fig_sp1014} we present a  selection of the  spectra acquired since 2010 in the range 4500-4800 \AA, which contains  the most diagnostic lines for spectral typing.
  In Table~\ref{tbl_spevol}  we report the spectral types   together with the corresponding  V magnitudes and the equivalent widths of  some H and He lines.
 
\begin{figure}
\epsscale{.90}
\plotone{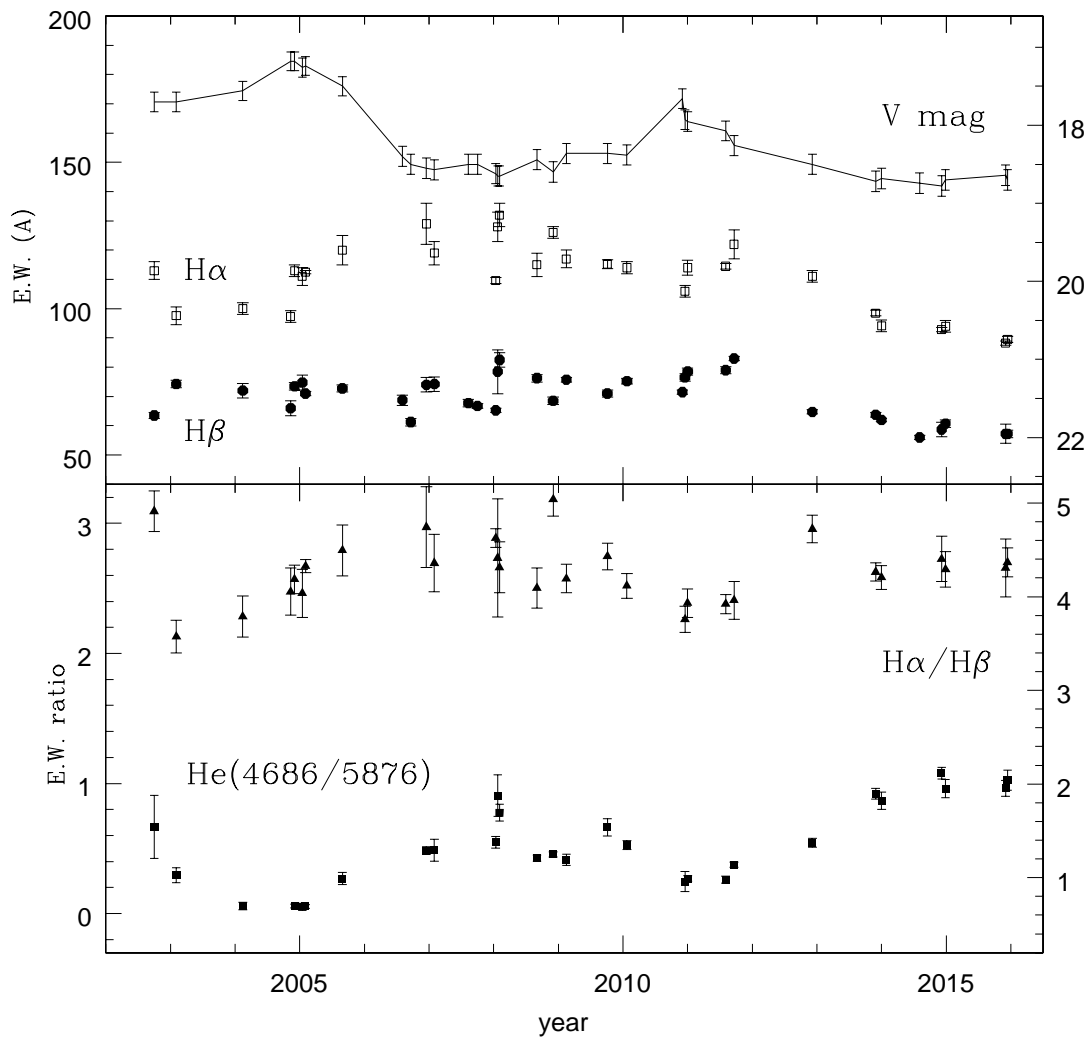}
\caption{  
The equivalent width (in ~\AA~)  and ratios of the main emission lines in the spectrum of GR~290 during 2002-2015. Top panel: e.w of H$\alpha$ (open squares) and of H$\beta$ (filled circles, multiplied by 2.5).
The  corresponding V magnitude  is shown for comparison ( connected points, left scale). In this and in the following plots only the EW derived from the Loiano and Cima Ekar Observatories have been plotted for homogeneity.
 Lower panel:  ratio of the equivalent width of He II 4686~\AA/He I 5876~\AA ( filled squares, left scale) and H$\alpha$/H$\beta$  ( filled triangles, right scale).   }
 \label{fig_EW_rap}
\end{figure}

The spectral evolution of the star during the last years is  mainly described by the changes in the equivalent width of  helium lines and of the broad blend of emission lines around 4650~\AA ~(the so called f-feature).
  Neutral and ionised helium lines have an opposite  behavior.
 The  hydrogen Balmer lines, H$\alpha$ and H$\beta$ are irregularly variable, but their variations seem not to follow the light curve, except for a probable gradual weakening in the last years. In particular, during the 2003-2014 period the H$\alpha$/H$\beta$ equivalent width ratio did not significantly vary around a mean value of 4.2.
 Figure \ref{fig_EW_rap} illustrates the time evolution of the hydrogen emission lines top panel)
  and the ratio H$\alpha$/H$\beta$ (lower panel) during the 2002-2015 period.  The lower panel  shows  also  the equivalent width ratio
  (\ion{He}{2}\,4686)/(\ion{He}{1}\,5876) which is smaller when the stellar apparent luminosity is higher. Furthermore, the apparently dumped EW oscillation of both the \ion{He}{2} 4686~\AA~ line and the 4686/5876 ratio might suggest a trend of increasing 
temperature.

\begin{figure}
\epsscale{.70}
\plotone{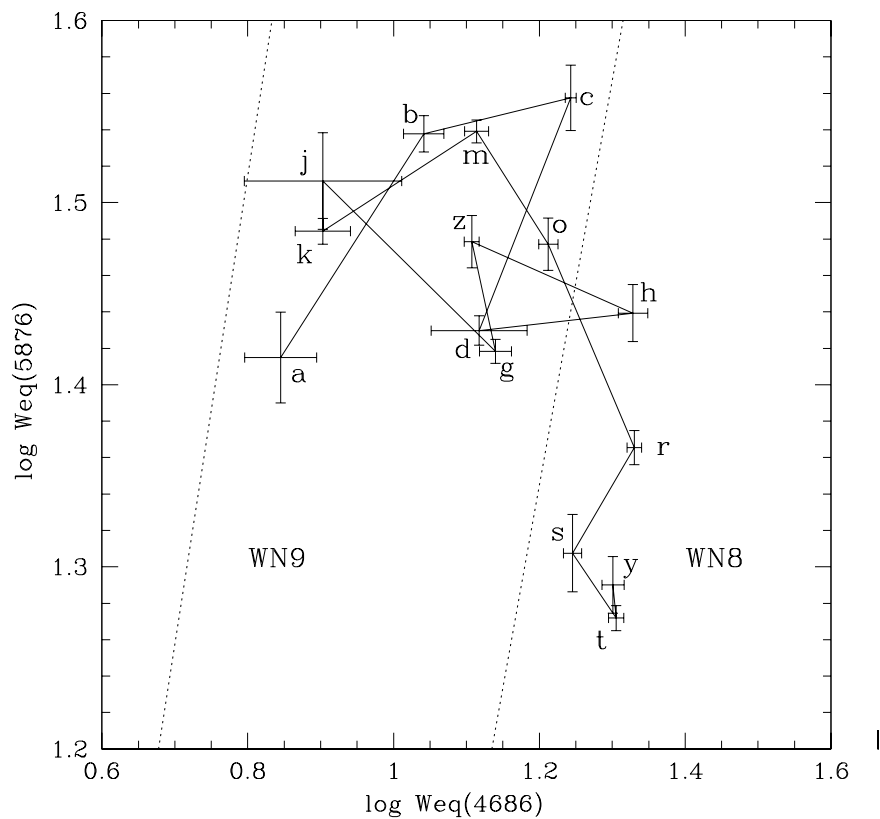}
\caption{The time evolution  of the equivalent width of He II 4686~\AA ~and He I 5876~\AA
 ~to mark the path of GR 290 through the WN sub-classes during its luminosity variations from August 2005 to December 2015.
  The letters correspond to the following dates: a=2005 Aug. 31;  b=2005 Nov. 08;  c=2006 Dec. 14; d=2007 Jan. 29; h=2008 Feb. 07; z=2008 Sep. 08; g= 2010 Jan. 21; j=1010 Dec.18; k=2011 Aug. 03; m=2011 Sep. 20; o=2012 Dec. 09; r=2013 Nov. 27;  s=2014 Jan 01; t=2014 Dec. 11; y=2015 Dec.11.
   \label{fig_Crow}}
\end{figure}

 The \ion{He}{2} 4686~\AA~ and \ion{He}{1} 5876~\AA~ lines have been used by  \citet{crow97} to distinguish stars with different spectral types. \citet{polc11} used the  equivalent widths diagrams EW(5876) vs EW(4686) to mark the path of GR~290 through the WN sub-classes during its recent luminosity variations.  
Since January 2010 the star has moved in the Crowther \& Smith diagram towards the WN8 locus and from December 2013 to December 2015 its representative point is well inside the WN8 locus, as it was at the beginning of 2008 during the previous minimum
(see  Figure~\ref{fig_Crow}).   
 
\begin{figure}
\epsscale{.99}
\plotone{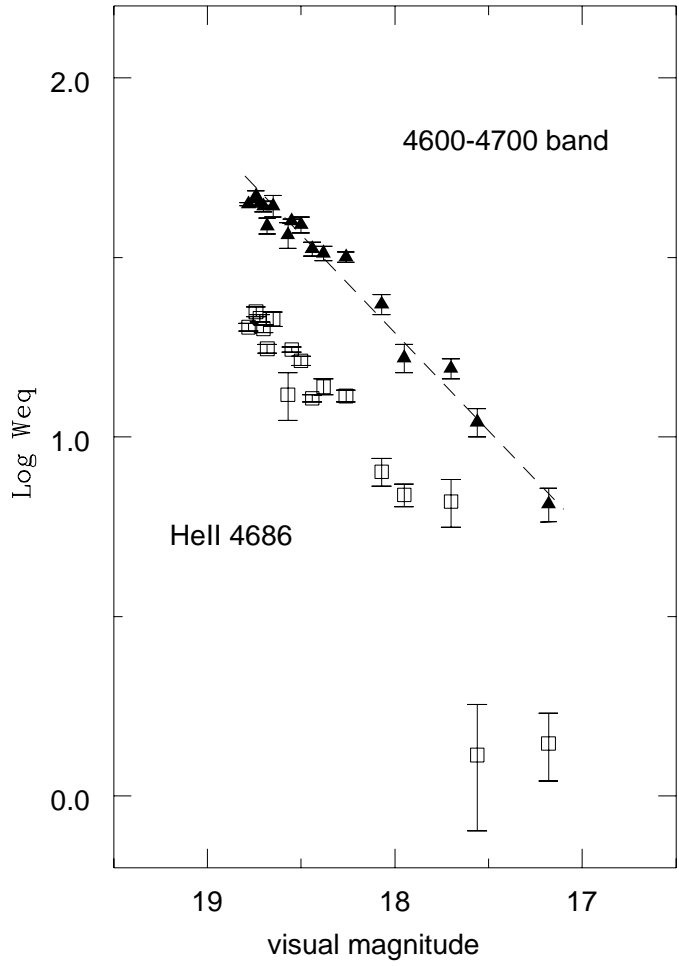}
\caption{ 
Logarithm of the equivalent width of the \ion{He}{2} 4686~\AA~ line (open squares) and the 4600-4700~\AA~ emission feature (triangles) versus visual luminosity during January 2003-December 2015.
The fit of the data points described  in the text is shown}
 \label{fig_blend}
\end{figure}

   The growing excitation temperature of WR stars from late  to earlier subtypes is best described by the intensity of the emission lines between 4600 and 4700 \AA. 
   The blend forming the  f-feature  as well as the  \ion{He}{2} 4686~\AA~ emission become prominent when the star is fainter as  was  ascertained by  \citet{polc11}  and shown   in Figure ~\ref{fig_sp1014}.
In order to study quantitatively this behavior, we have measured the equivalent width of the emission in the 4600-4700~\AA~ range avoiding the lower excitation ~\ion{He}{1} 4713~\AA~ line which is too strong in the later WN spectral types. In Figure~\ref{fig_blend} we plot the EW (in logarithm) of both the \ion{He}{2} 4686~\AA~ line and the 4600-4700~\AA~ blend as a function of the $V$~ magnitude.  
The 4600-4700 \AA ~blend presents a marked linear correlation with
the V magnitude  in an ample luminosity range with a mean slope of 0.544 $\pm$ 0.027.
We remark that this correlation does not vary, within the errors, during three different optical variability  cycles.
For the HeII line the linearity fails when the star is bright, possibly because of the large uncertainty on the lowest  equivalent widths.

\section {Physical parameters of GR 290} \label {parameters} 

\citet{mary12a} and \citet{clark12}, using the {\sc cmfgen} atmospheric modeling code \citep{hillier98}
have derived the physical parameters of GR~290 using spectra taken in February 2005 (V=17.2) and October 2007-January 2008(V=18.6), and in September 2010 (V=17.9). 
Presently, we have at our disposal many spectra of the star collected mainly at the Loiano Observatory and at the Special Astrophysical Observatory   covering the period of October 2002 -- December 2014 when the star displayed an ample range of variation in visual luminosity.
 This time interval covers two light maxima and three minima. In order to see how the parameters of GR~290 changed with time we 
have modelled the most representative spectra with best quality, obtained during this period.

  We have  applied the same procedure as \citet{mary12a} and have constructed nine models. Details on the procedure may be found in \citet{mary12a} and are not repeated here except where necessary.

Every model is defined by a hydrostatic stellar radius R$_*$, luminosity L$_*$, mass-loss rate $\dot{M}$, filling factor $f$, wind terminal velocity $v_\infty$, stellar mass M$_*$ and by the abundances Z$_i$ of the element species. 

We started modelling by fitting the spectra obtained in  January 2005 during a phase of maximum brightness and in December 2008. These spectra are most suitable as  having the widest spectral range and the best resolution among our spectra.    
    Then for every spectrum  we have calculated few models  giving a similar fit with slightly different mass loss rate, velocity, temperature,  radius, filling factor.
 The uncertainties reported below include the range of parameters of the models. We assumed that element abundances do not change with changing phases. Thus we used the abundances derived for the December 2008  spectrum for fitting  all the remaining data. 

Various attempts to derive the H/He abundance ratio gave a large range of possible values from 1.5 to 1.9;     differences between the  models are not evident  unless all the others parameters are fixed.    
Ultimately  we decided to assign a  value of 1.7 to the H/He ratio.
We estimated the abundances of carbon (N(C)/N(He) = $1^{+0.4}_{-0.3}\times10^{-4}$) and nitrogen (N(N)/N(He) =$3^{+1.5}_{-1}\times10^{-3}$) by fitting the WR blue bump of the blend at 4630-4713~\AA.
 The determination of the oxygen fraction is a more difficult task since there are no oxygen emission lines in the spectrum of GR~290. We, however, noticed that the intensity of the \ion{He}{2} lines decreases when we include oxygen in our model. Thus we estimated the oxygen abundance indirectly by fitting the \ion{He}{2} lines and obtained N(O)/N(He) = $2^{+2}_{-1}\times10^{-4}$. For the abundances of Ar, S and Ne we used those of M~33 from \citet{magri10}. For the abundances of other elements (Mg, Al, Si, Ca and Fe) we assumed a value of half solar as the metallicity of M~33 is low. 
We compared several models and concluded that the abundances of these elements do not affect model spectra significantly.

{\sc cmfgen} allows to build the models with different velocity laws describing the radial dependence of the wind velocity. Often the same spectrum may be equally well described by either simple $\beta$ law or more complex double-$\beta$ profile, yielding different values of luminosity and temperature. 
In this work our main task is to track the changes of stellar parameters over time.
Due to the lack of theoretical works on this topic, it is difficult to say now which velocity profile is more physically justified; this  is a question we leave for a separate publication. Here  we found sufficient  to use the same simple wind velocity law with $\beta=1$ for all  models. 
Due to fact that the profiles of spectral lines are very sensitive to the value of the terminal velocity, we can estimate $v_{\infty}$ using even low resolution spectra. 
At minimum  brightness we used the P Cyg profiles of  the He I triplet lines (such as $\lambda\lambda$~3889 \AA, 4025 \AA, 4471 \AA and 6678 \AA) and derived   V$_{\infty}$ = 400 km$s^{-1}$.
 At maximum brightness  we used the emission lines fits and found   V$_{\infty}$ = 200-250 km$s^{-1}$. Thus
  V$_{\infty}$  changes from the hot phase to the cool phase, and this changing is real.
  Indeed, if at maximum brightness the velocity were equal to 400 km$s^{-1}$, then we would have detected P Cyg profiles even in our low resolution spectra.  However  P Cyg profiles are not observed at maximum brightness  in the spectrum obtained in January 2005 with spectral resolution $\Delta \lambda=3.5$ \AA. 

 For measuring $T_{\rm{eff}}$ with {\sc cmfgen}  we compared  the  intensities of different ion lines (He{\scriptsize I},{\scriptsize II}; C{\scriptsize III},{\scriptsize IV}; N{\scriptsize II},{\scriptsize III}; Si{\scriptsize III},{\scriptsize IV}), i.e we used the traditional ionization-balance method.
For the C ionization structure, we used the C~III $\lambda\lambda$ 4647-50 \AA\ diagnostic lines. For the N:  N~II 4601-07-30-43 \AA,  N~III 4510-15-18-24-31-35-47 \AA, N~III 4634-40 \AA.
The Si ionization structure was constrained through the relative strength of  Si~III, and Si~IV lines, using lines of Si~III 4553-4568-4575~\AA ~and Si~IV 4088-4116 \AA.
By cross-checking all these indicators we estimated an uncertainty of  $\pm$~1.000~K.

\medskip

 To  estimate  the luminosity of the object, hence the stellar radius, the magnitudes of the star were calculated in the $B$ and $V$ filters from the model spectra and compared with observations.
The  model $(B-V)_0$ is determined by the  temperature, which was estimated as described above;
 this model color is then  compared with  the observed $B-V$  from which comparison the  reddening is determined.

 Result is that the  model $(B-V)_0$ is consistent with observed B-V only if E$(B-V)<$0.08, assuming
  for M~33 a distance of D=847$\pm$60 kpc \citep{galleti04}.

Actually the  Galaxy foreground extinction is $ E(B - V) = 0.052$ (according to the NED extinction calculator, Schlegel et al. 1998) and the intrinsic extinction in  M~33 is negligible according to the detailed dust maps  by \citet{hippe03}.

Furthermore E$(B-V)$  have also been  computed for four close stars whose magnitudes and spectral types are given in
 Massey et al. (2006).
The closest star, a B0.5Ib, has E(B-V)=0.06, nearly the same as found for GR290.

Finally we have computed models with  extinction up to  E$(B-V)$ = 0.15, as assumed by \citet{hum14}.
Of course higher extinction would imply a significant increase of the  luminosity.
However, in all the minimum phases GR290  would lie  still outside  the instability strip of  LBV stars.

\medskip

The mass-loss rate was obtained by reproducing the intensities of the strongest lines --  namely, those of hydrogen, helium, and nitrogen. 
 
The parameters of GR~290  were also obtained by \citet{mary12a}  during the maximum brightness in 2004-2005. 
For this goal these authors used a spectrum obtained in February 2005 at  the Russian 6-m telescope with spectrograph SCORPIO with spectral resolution $\Delta\lambda=10$ \AA. 
 To reproduce that spectrum a  volume-filling factor  $f$=0.5 was necessary.
 \citet{clark12} modeled the spectrum of Romano's star obtained in September 2010 when the $V$-band magnitude of the object was between 17.75 and 17.85. They also used the {\sc cmfgen} code;  the best fit was obtained with  $f$=0.25.
Since in the current work we used the spectrum obtained in January 2005   with better resolution  ($\Delta\lambda=3.5$ \AA), we tried to fit the observed spectra with  several values of the filling factor.
 We found that already  models with $f$ =0.1 well  describe all spectra;
$f$=0.2 equally gives good results. The other parameters change very little with respect to   the first  case, the main difference,  being in T$_{eff}$,  is always within the errors.
 Models with $f$=0.3  give a worse fit for  every other parameters.
 From the comparison of observed with model spectra it is not clear which value of $f$ is more correct.  We assumed as final value $f$=0.15.
  
\begin{figure}
\epsscale{.99}
\plotone{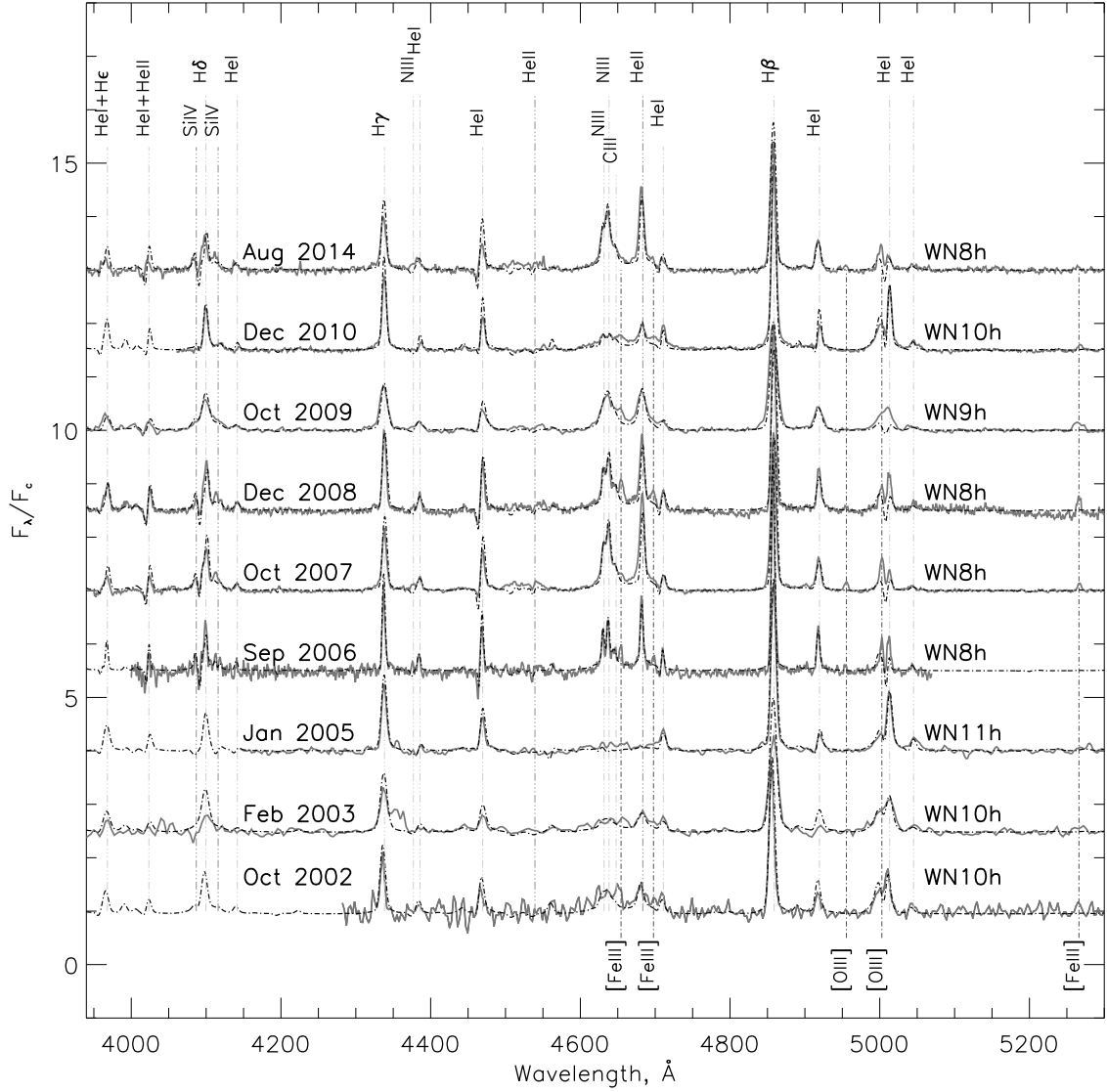}
\caption{  Normalized optical spectra of GR~290 compared with the best-fit CMFGEN models (dash-dotted line). The model spectra are convolved with a Gaussian instrumental profile.
       \label{fig_spmod}}
\end{figure}

 Table~\ref{tbl_logfile} lists all spectroscopic observations of GR~290 used to construct the models.
Details on the instrumental characteristics are described in Paper 1 and in \citet{mary12a}.
The observed spectra and the models  themselves are shown in Figure~\ref{fig_spmod}.
Table~\ref{tbl_parmodel} lists the parameters of these models.

The accuracy on the derived estimates of luminosity and radius depends on several parameters, including the uncertainty on the distance and on the interstellar extinction. 
However, since the main scope of our work is to estimate the amplitude of the variation of the parameters and its accuracy, in Table~\ref{tbl_parmodel} we only give the errors derived from the fitting procedure.
It is worth noting that at maximum brightness He{\scriptsize II} and  N{\scriptsize III} lines are absent (or very weak), thus our estimations give effective temperature in the range 20-23\,kKš. This range of values of $T_{eff}$ leads  to a large scatter in the luminosities, which  at maximum brightness results  in the range $0.75-1 \cdot 10^6 L_{\odot}$.

\begin{figure}
\epsscale{.80}
\plotone{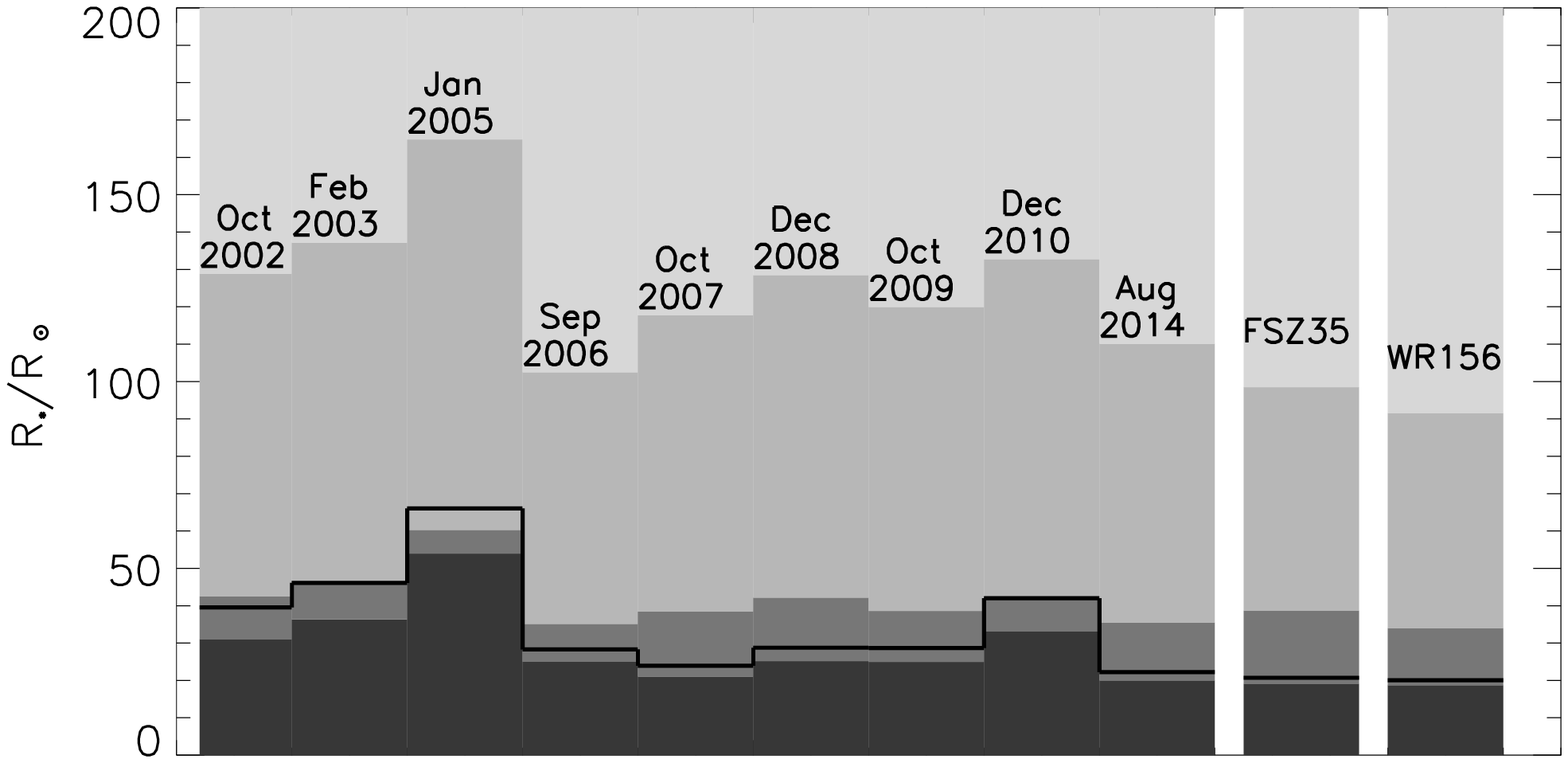}
\plotone{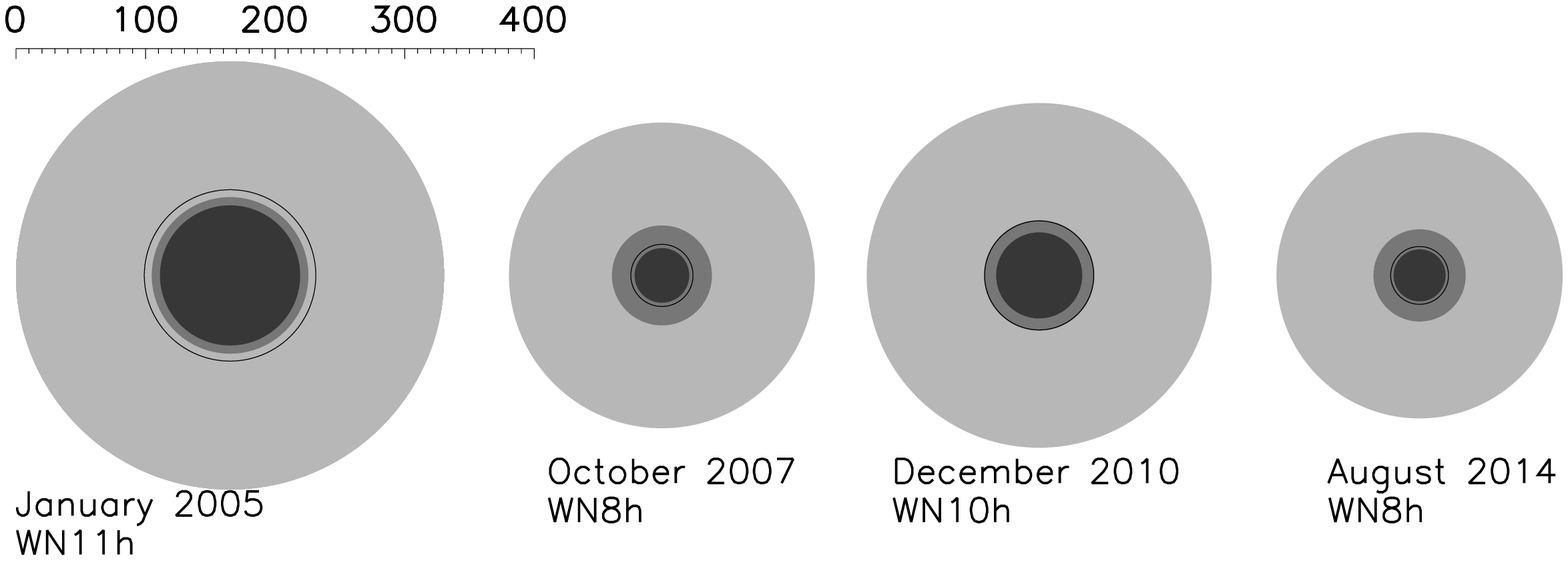}
\caption{ Change of the wind extent with time. Regions with n$_{e}>$10$^{12}$ cm$^{-3}$ are shown in black, 10$^{12}>$n$_{e}>$10$^{11}$ cm$^{-3}$ in dark grey, 10$^{11}>$n$_{e}>$10$^{10}$ cm$^{-3}$ in grey, 10$^{10}>$n$_e$ cm$^{-3}$ in light grey. The solid black line shows the radius where the Rosseland optical depth ($\tau$) is 2/3. WN8h stars FSZ35 and WR156 are also shown for comparison.
 \label{fig_wind}}
\end{figure}

We can see from Table~\ref{tbl_parmodel} that the radius $R_{2/3}$, where the Rosseland optical depth is equal to 2/3, is as low as 22-26~$R_\sun$ near the minima and increases up to 63~$R_\sun$ during the 2005 maximum. On the other hand, the temperature at $R_{2/3}$ decreases from 31-33~kK (with WN8-9h spectral type) to 25~kK (WN11h). The nature of the stellar wind significantly changes, being much denser and slower during the eruption in 2005. Figure~\ref{fig_wind} illustrates the structure of the wind-filled envelope of GR~290 during different luminosity phases. Note that the $R_{2/3}$ surface moves deeper in the wind during the minima. For comparison  Figure~\ref{fig_wind}  shows the structure of the wind of two WN8h stars: FSZ35 \citep{mary12b} and WR156 \citep{mary13b}, belonging to M~33 and our Galaxy, respectively. It is clear that the wind structure of GR~290 during the minimum of brightness is fairly similar to the one of typical WN8h (non variable) stars.

Our estimates are consistent with those by \citet{clark12} within the errors, with the exception for the luminosity. It must be noted that \citet{clark12} assumed a distance to M~33 of 964~kpc, while we adopted a distance of 847~kpc. Due to this difference in the assumed distances, we cannot directly compare the model luminosities, as the luminosity influences both the apparent magnitude of the star and the mass loss rate, which is crucial for all other parameters.

\begin{figure}
\epsscale{.99}
\plotone{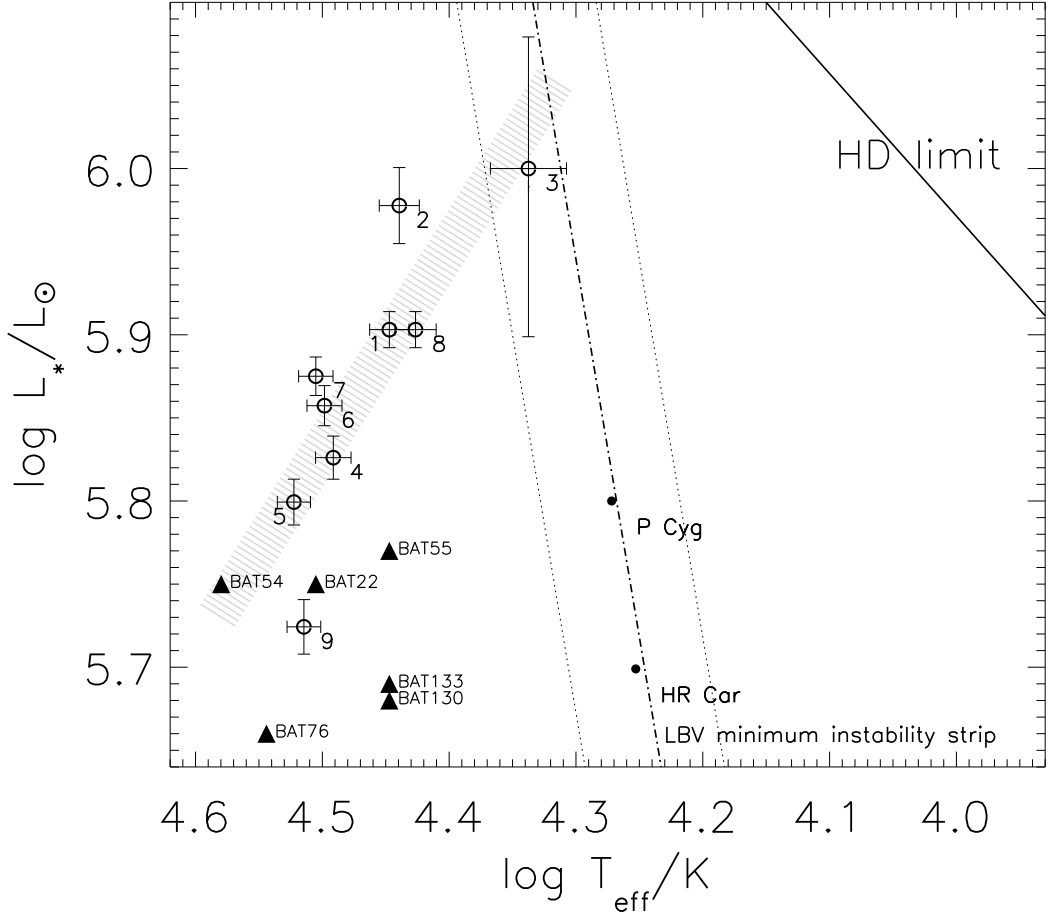} 
\caption{The position of GR~290 in the H-R diagram. The LBV minimum instability strip is traced with a dashed-dotted line, the Humphreys-Davidson limit (Humphreys \& Davidson, 1994) is shown with a solid line. 1~Oct 2002, 2~Feb 2003, 3~Jan 2005, 4~Sep 2006, 5~Oct 2007, 6~Dec 2008, 7~Oct 2009, 8~Dec 2010, 9~Aug 2014. Also the LBV stars  P Cygni and HR\,Car are shown with circles. Data for these objects were taken from   Najarro (2001) and Groh et al. (2009b). Triangles mark late-WN stars taken from Hainich et al.2014.  The hatched strip shows the region crossed by GR 290 during its recent luminosity  cycles.
       \label{fig_HR}}
\end{figure}

The main result of this analysis is that the bolometric luminosity of GR~290 is variable, being higher during the phases of greater optical brightness. Previously, \citet{polc11} found a significant decrease in the bolometric luminosity between the 2005 maximum and the 2008 minimum. 
Later, \citet{mary12a} gave further proofs that the bolometric luminosity of GR~290 has really decreased by about 40\%. The present model fitting of a large sample of spectra obtained during two successive luminosity cycles definitely confirm their result and allows to trace the recent path of the star in the Hertzsprung-Russell (H-R) diagram shown in Figure~\ref{fig_HR}.

As a further check of the above result, we tried to construct models with constant bolometric luminosity, but the resulting model stellar magnitudes differ from the observed ones.
For example, if we take $L_*=6\cdot 10^5 L{\scriptsize \sun}$~=const then the difference between the December~2008 model stellar magnitude and the observed one is about 0.2 mag. Such difference might be due to photometric  errors. But for the February~2005 model such difference exceeds 0.7 mag, and this can not be in any way attributed to photometric uncertainties. Therefore we conclude that the bolometric luminosity variation is real. 

\section{The evolutionary state of GR 290} \label{evolution}

\subsection{GR~290 and the LBV stars} \label{GR290-LBV}

  As discussed in the previous sections, GR 290, from the observational point of view,
shares some properties of the strongly active LBVs as classified by \citet{vang01}.
However, GR 290 is different from all other known LBVs for being at any
luminosity phase much hotter, since it presently exhibits a WN9-11h spectral type during brightness
maxima and a WN8h spectral type at the minima. Also the spectral type
shown by GR 290 during the 1992-94 absolute maximum (equivalent to B, probably late
spectral type) is hotter than the ones of the majority of other LBVs, that usually have A phases.
or even F spectral types during their maximum.
 Furthermore, if the present 
correlation between the visual magnitude and the spectral
type had been  valid also before 1960, GR 290 (being its luminosity weaker than B=17.5 mag in
all the recorded data) would have been much hotter than all the so far known dormant LBVs (all having an Ofpe - or WN 9-11 - spectral type)
also during its long lasting quasi-quiescent phase in the first half of the 20th Century. 

Last, all the stars so far considered to be dormant LBVs (but one: HDE326823,
that was suggested to be in a post-LBV phase by Sterken et al. 1995) are placed near to
the lower boundary of the instability strip \citep{wolf89}. 
On the contrary, the GR 290 representative point in the H-R diagram, as seen in Figure~\ref{fig_HR}, presently
is far from the instability strip. 

We also stress that the spectral and luminosity variations of GR 290 in the last years occurred at nearly
constant $(B - V)$ color index, contrary to what expected for an LBV. 

It is thus unavoidable to conclude that presently there are no known LBV stars that can be
directly compared with GR 290 in every aspect.  

\subsection{GR~290 and the WNL stars} \label{GR290-WNL}

\begin{figure}
\epsscale{.99}
\plotone{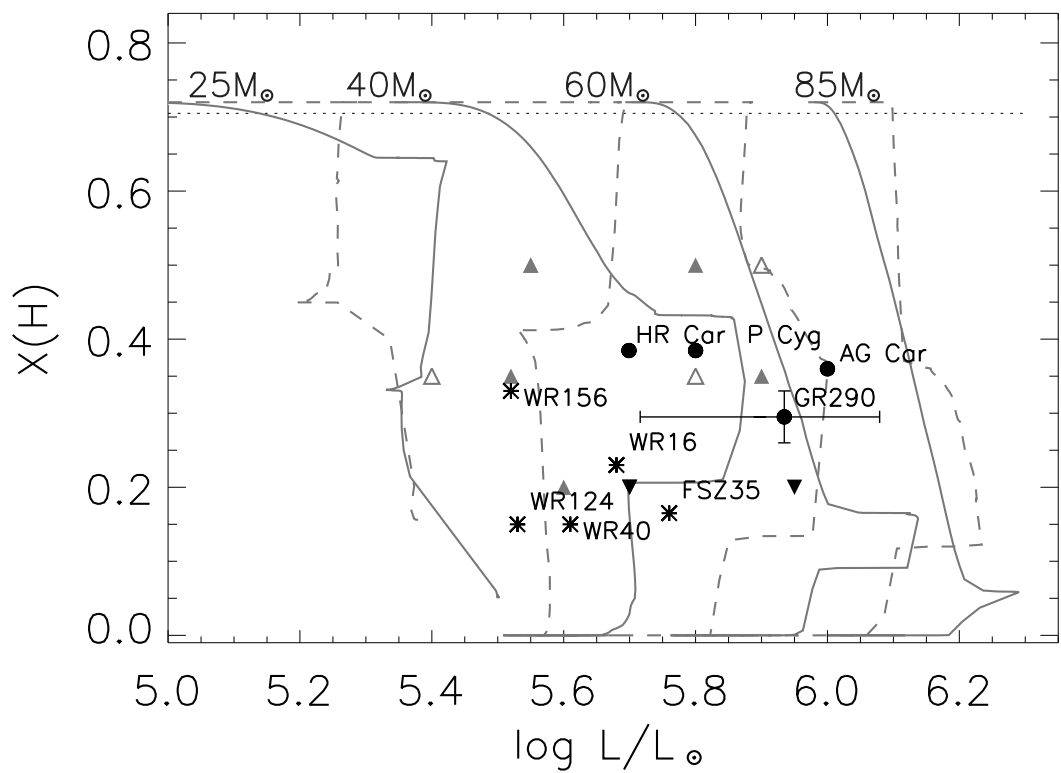}
\caption{ 
Hydrogen mass-fraction as a function of luminosity.
Continuous lines show the Geneva tracks for non rotating models.
The position of the LBV and WN stars is also marked. Late WN stars in M31 are shown by triangles, data taken from Sanders et al. (2014) . WN9 stars are represented with open triangles, WN8 with filled triangles and WN7 with upturned triangles. 
 \label{fig_Hmf}}
\end{figure}

A point to be considered is the present high temperature of GR~290 and its spectrum at minimum very similar to that of WN8h stars. Figure~\ref{fig_HR} shows that the luminosity and temperature of GR~290 fit well that of late WN stars. The mass loss rate of GR~290 is also similar, but the wind velocity of GR~290, $v_\infty$ = 400 km~s$^{-1}$, is significantly lower than that of WN8h stars (see Maryeva \& Abolmasov 2012a,b). Also the hydrogen abundance of the envelope appears higher than in WNL stars. 
Figure~\ref{fig_Hmf} shows the diagram of the hydrogen mass-fraction as a function of luminosity for different evolutionary tracks. GR~290 lies higher than WN8h stars. These results give evidence that, from the evolutionary and structural point of view, GR~290 is younger than WN8h stars. 

\subsection{Evolutionary considerations} \label{considerations}

According to the above derived bolometric luminosity, ($\log L_*/L_{\odot}$ =5.8--5.9) and effective temperature 
of 22$-$33 \,  kK
the position of GR~290 fits the 60 $M${\scriptsize $\sun$} evolutionary tracks of \citet{chieffi13} with and without rotation,
 just to the left
 of the low temperature loop in the H-R diagram. The Geneva evolutionary tracks for non rotating models suggest a similar initial mass (see e.g. \citet{groh13}). Although all these tracks were computed for solar metallicity, the initial mass of GR~290 cannot be too far from this value.  According to the evolutionary models of \citet{chieffi13} GR290 should now be 4 Myr old and have a mass of 26 $M${\scriptsize $\sun$}. 
The star is thus in the mass range where theoretical evolutionary models foresee that, after the O phase and before the Hydrogen-poor Wolf-Rayet (WR) star phase, stars transit from an Of/WN  (or WN 9-11) through the LBV phase to a WNL phase. 
\cite{groh14} computed the spectral evolution of a non-rotating 60 M{\scriptsize $\sun$} star, that could be appropriate to the case of GR~290. They found that the star first evolves to the right part of the H-R diagram through the hot LBV phase (near the end of core H-burning) to the cool LBV loop of the evolutionary track. Then, after the beginning of core He burning, the star leaves the LBV phase and evolves to early WN, rapidly crossing in a few thousand years the late-WN phase. 
In the Groh et al. framework, the effective temperature and spectral type of GR 290 would clearly place the star in the WNL 
phase. 
The post LBV phase is further supported by the helium and nitrogen abundances derived in Section \ref{parameters}, which are higher than in the classical LBVs. One should also note that, according to \citet 
{groh14}, the lifetime of a 60 M{\scriptsize $\sun$} star in the LBV phase
 is 50 times longer than that in the post-LBV WNL phase, hence we should expect WNL stars to be very rare. 
The question is whether in the post LBV phase stars may still undergo large luminosity variation such as those observed in GR~290.  

\subsection{Luminosity variation} \label{L-variation}

As discussed in Section \ref{parameters}, a distinguishing peculiarity of GR~290 is the variation of its bolometric luminosity during its recent apparent  luminosity  cycles. The evolution of LBVs during the S\,Dor cycles seems to occur in most cases roughly at constant bolometric luminosity (see, e.g., 
\citet{smi94}; \citet{dekoter96}; \citet{walb08}). Indeed, the variation of the bolometric luminosity is a point so far observed  only in a few LBV cases: \citet{groh09a} have found that the bolometric luminosity of AG Car has decreased by a factor of 1.5 from minimum towards the light maximum of the S\,Dor cycle. A reduction of the bolometric luminosity during the cycle was also derived by \citet{lamers95} for S\,Dor itself. \citet{lamers95} interpreted this result in terms of the radiative power being partially transformed into mechanical power in order to expand the outer layers of the star from minimum to maximum. The same explanation was given by \citet{groh09a} in the case of AG Car and by \citet{andr78} to explain 
 the one magnitude decrease of the bolometric luminosity of $\eta$ Car after the 1843 giant outburst. 
In the case of GR~290 the power associated to its mass loss rate is by far too low to explain its changes in bolometric luminosity. In addition, contrary to the cases discussed above, the L$_{bol}$ increases of GR~290 are recorded during its visual luminosity  maxima. 
If, as for instance suggested by \citet{guzik14}, the GR~290 active phase were triggered by bursts of nuclear energy, an associated increase in bolometric luminosity could well occur at the light maxima as observed,
due to the energy release associated with these nuclear runaway events.
Since changes of the bolometric luminosity of such an amplitude have not been so far recorded in any LBVs we may argue that the higher temperature of GR 290 plays an important role in the above suggested process.

\section{Conclusions} 
In summary, we have shown that:

1) During the last one hundred years GR~290 experienced two phases: one long lasting quasi-stationary, presumably hot phase. Then an active phase started in mid-20th Century with five luminosity maxima separated by 7-10 years. 

2) The amplitude of the last variability cycles is decreasing with an associated increase of the stellar temperature at minimum. 
During this period of time the stellar temperature is modulated in anti-correlation with the visual luminosity. 

3) The star is in all phases much hotter than the other LBVs, reaching an effective temperature of 33 kK at the apparent luminosity minima. 

4) During the last two variability cycles the bolometric luminosity changed following the variation of visual luminosity, being a factor ~1.5 higher at maximum.  

5) The bolometric luminosity of GR~290 fits 
the evolutionary tracks of a $\simeq$ 60 $M\Sun$ star. 
The high effective temperature and WNL-type spectrum place the star 
after the low temperature loop of the evolutionary tracks which is thought to be occupied by the LBV stars. 

6) At present, the representative point of GR~290 in the H-R diagram is close to that of WN8h stars, but the chemical abundances suggest the star 
be younger than them.

The above discussed points lead us to conclude that GR~290 is presently in a short, and thus very rare, transition phase between the LBV evolutionary phase and the nitrogen rich W-R stellar class, as also suggested by \citet{polc11} and \citet{hum14}.

GR~290, though is no longer near the LBV minimum instability strip, still might be subject to structural instabilities similar to those which gave rise to the last observed large variability cycles.

We cannot foresee how long these instabilities will last. Following \citet{guzik14} the instability is due to the hydrogen mixing in the core generating a burst of nuclear energy: presently, such repeated nuclear events should be the extra-energy input to increase the bolometric luminosity of GR~290 during its last outbursts. 
They will end when the hydrogen percentage in the envelope will become too low to be mixed in the core.  
However, the steady decreasing amplitude of the last cycles and  gradual increasing of the temperature of GR 290 both suggest that this instability phase  will not last too long.
This point will require new theoretical studies on the internal instabilities in luminous stars and GR~290 could probably be an ideal target to study this process.

\clearpage

\begin{deluxetable}{lcclrccrcl}
\tabletypesize{\footnotesize}
\tablewidth{0pt}
\tablecaption{Photometry of GR~290 from archival plates.\label{tbl_archmag}}
\tablehead{
 \colhead{Plate no.}    & \colhead{date}  & \colhead{JD}  & \colhead{emulsion}  & \colhead{exp}  & 
\colhead{rms}  & \colhead{slope}  & \colhead{mag}  & \colhead{band}   & \colhead{remarks}     }
\startdata
       &         &       &      &      (1)    &      (2)    &     (3)   &       &     & (4)  \\
Hamburg    &         &       &      &         &         &       &       &     &              \\
   S00332  &   1914 09 21 & 2420397 & Hauff-Ultra        &    30  &  0.08  & 1.03 &   18.4  &  B  &  poor   \\  
   S02358  &   1920 10 18 & 2422616 & unknown            &    94  &  0.07  & 0.99 &   18.2  &  B  &  good   \\  
   S02396  &   1920 11 12 & 2422641 & Agfa-Isorapid      &    40  &  0.12  & 0.95 &   18.4  &  B  &  poor   \\  
   S02436  &   1921 02 09 & 2422730 & Agfa-Isorapid      &    40  &  0.18  & 1.03 &   17.9  &  B  &  poor   \\  
   S02587  &   1921 09 09 & 2422942 & Hauff-Ultra        &   114  &  0.13  & 1.11 &   18.5  &  B  &  fair   \\  
   S02676  &   1921 11 24 & 2423018 & Agfa-Isorapid      &    60  &  0.12  & 0.92 &   18.5  &  B  &  fair    \\ 
   S02982  &   1922 11 18 & 2423377 & Agfa-Isorapid      &    30  &  0.12  & 1.02 &   18.4  &  B  &  fair    \\ 
   S04042  &   1928 11 09 & 2425560 & Agfa-Astro-Inhalo  &    60  &  0.10  & 1.13 &   18.4  &  B  &  good    \\ 
   S04295  &   1930 08 24 & 2425483 & Agfa-Astro-Inhalo  &    47  &  0.08  & 1.09 &   18.4  &  B  &  good    \\ 
Pulkovo     &         &       &      &         &         &       &       &     &              \\
   K235   &     1935 11 16 & 2428123 & unknown   &      60  &  0.05  & 0.78  & $>$ 16.5     & B & not visible  \\
   K397   &     1937 09 30 & 2428807 & unknown   &     100  &  0.28  & 1.20  &   18.67   & B & faint	    \\
   K412   &     1937 10 07 & 2428814 & unknown   &      90  &  0.20  & 1.02  &   18.45   & B & faint	    \\
   K413   &     1937 10 07 & 2428814 & unknown   &     105  &  0.08  & 0.84  &   17.95   & B & faint	    \\
   K508   &     1938 11 27 & 2429230 & unknown   &      66  &  0.12  & 0.84  &   17.93:: & B & very faint   \\
   K510   &     1938 12 17 & 2429250 & unknown   &      80  &  0.08  & 1.09  &   18.54   & B & very faint  \\
Heidelberg   &         &       &      &         &         &       &       &     &              \\
  D0154   &    1907 08 08 & 2417795  &unknown  &        35  &  0.10  & 1.31 &     18.19 &  B & faint         \\
  D0169   &    1908 08 12 & 2417799  &unknown  &       124  &  0.10  & 1.31 &  $>$ 17.5   &  B & not visible   \\
  D0359   &    1908 07 25 & 2418165  &unknown  &        40  &  0.10  & 1.98 &     18.85 &  B & faint	     \\
  D1615   &    1917 12 10 & 2421572  &unknown  &        40  &  0.15  & 1.00 &     17.79 &  B & faint 	     \\
  D1616   &    1917 12 17 & 2421579  &unknown  &        40  &  0.21  & 1.04 &     17.97 &  B & faint           \\
  D1617   &    1917 12 17 & 2421579  &unknown  &        40  &  0.11  & 1.04 &  $>$ 17.5   &  B & not visible     \\
  D1631   &    1918 01 30 & 2421623  &unknown  &        40  &  0.13  & 1.35 &   $>$ 17.5   &  B & not visible     \\
  D1632   &    1918 01 30 & 2421623  &unknown  &        40  &  0.20  & 1.54 &  $>$ 17.5   &  B & not visible     \\
  D2118   &    1920 08 15 & 2422551  &unknown  &        60  &  0.19  & 1.46 &  $>$ 17.5   &  B & not visible     \\
  D2119   &    1920 08 15 & 2422551  &unknown  &        70   & 0.20  & 1.42  & $>$ 17.5    & B  &not visible     \\
  D2485   &    1922 10 14 & 2423341  &unknown  &        90   & 0.12  & 1.02  &    17.90  & B  &visible	       \\
  D2498   &    1922 10 25 & 2423352  &unknown  &        64   & 0.20  & 1.48  &    18.40  & B  &faint	       \\
  Yerkes   &         &       &      &         &         &       &       &     &              \\      
Ry-8:   &  1901 08 15  &   2415613 & unknown   & 210 & 0.14 & & 18.33 &  B & \\
scanMP  & 1902 09 04  & 2415999 &  unknown    & 240 &  0.10 & & 18.50  & B &  \\ 
G107a   & 1910 08 06   & 2418890 &  unknown    & 510 &  0.10 & & 18.26  & B & 60" Mt Wilson \\ 
R-3481:  & 1916 09 19   & 2421127 &  unknown   & 50 &  & &  $>$17.95 & B &  not visible \\  
R-3518:  & 1916 12 09   &2421208   & unknown +W12  & 240 &  & &  $>$14.81 & V ? &  very dark  plate \\  
R-5478   & 1936 11 10  &2428484 &  I-O   & 30 & 0.5 & &  17.75 & B &  poor \\  
R-5523:  & 1937 01 10  & 2428545 &  I-C+ red   &60  &  & &  $>$16.95 & V-R &  poor \\  
R-5531:  & 1937 01 15  & 2428550 &  I-O   &37  &  & & $>$17.95 & B &  not visible \\  
R-5614:  & 1938 10 24  & 2429197 &  Agfa pan+yellow   & 60 &  & &  $>$14.95 &  V &  not visible \\  
R-5616:  & 1938 10 24  & 2429197 &    Agfa pan+yellow  &60  &  & &  $>$14.95 & V&  not visible \\  
R-5618:  & 1938 10 27   & 2429200  & E-40    & 5 &  & &  $>$15.43 & B &  not visible \\  
R-5621:  & 1938 10 27  & 2429200 & unknown    & 5 &  & &  $>$15.45 & B &  not visible \\  
R-5625:  & 1938 11 29   &  2429233 &  Agfa SPP+\#28   & 40 &  & &  $>$14.59 & R &  not visible \\  
cl 204  & 1947 09 13   &  2432442 &  unknown   & 402 &  & & $>$17.95 & B & not visible  \\  
vB9404 & 1952 10 12   &  2434299  &     & &  0.06 & & 18.40 & B &  \\  
Palomar    &         &       &      &         &         &       &       &     &              \\		      
  POSS-IE  &  1949 12 21 & 2433272 & 103aE+plexi   &  45  &  0.15  & 2.11 &  18.68  &   R  &   \\	       
  Quick-V  &  1982 10 18 & 2445261 & IIaD+W12      &  20  &  0.06  & 1.86 &  17.81  &   V  &   \\
  POSS-IIB &  1991 09 30 & 2448530 & IIIaJ+GG395   & 60   &  0.05  & 2.58 &  16.54  &  B   &   \\
  POSS-IIR &  1991 10 05 & 2448535 & IIIaF+RG610   & 75   &  0.06  & 2.66 &  16.04  &  R   &   \\
Asiago  &         &       &      &         &         &       &       &     &              \\   
 5624   &  1972 09 16  &2441577 & 103aO+GG13 &  30  & 0.18  &1.72  &   17.25  &  B & good \\
 5831   &  1972 11 07  &2441629 & 103aD+GG14 &  20  & 0.08  &----  &   17.43  &  V & good \\
 6804   &  1973 10 27  &2441983 & 103aO+GG13 &  20  & 0.21  &1.93  &   17.15  &  B & good \\
 6940   &  1973 12 15  &2442032 & 103aD+GG14 &  30  & 0.13  &----  &   17.83  &  V & good \\
11699   &  1982 10 14 &2445257 & 103aO clear     &  30  & 0.19  &1.78  &   16.70  &  B & high background  \\
12338   &  1983 11 28 &2445667&  103aD+GG14 &  30  & 0.12  &  &  $>$17.60  &  B & not visible \\
\enddata						       
\end{deluxetable}

\clearpage			

\begin{deluxetable}{cccccl}
\tabletypesize{\footnotesize}
\tablewidth{0pt}
\tablecaption{New photometric observations of GR~290.\label{tbl_newmag}}
\tablehead{
\colhead{Date} & \colhead{$B$}  & \colhead{$V$}  & \colhead{$R$}  & \colhead{$I$}  & \colhead{Obs}\tablenotemark{a}  }
\startdata
 2010 12 10     &- &       17.95 .10      &- &- &                   Greve \\
 2010 12 28     &- &        17.88 .07       &- &- &                 ARA  \\
 2011 01 03 &17.96 .05  &17.95 .05 & 17.73 .04 &- &         Loiano\\
 2011 01 05 &17.81 .05   & - & - & - &                                SSON \\
 2011 01 08       &- &     17.92 .05     &- &- &                     ARA \\
 2011 01 23       &- &     17.92 .07     &- &- &                  Greve \\
 2011 02 05       &- &     17.80 .10      &- &- &                    Greve\\ 
 2011 02 05       &- &     17.82 .06     &- &- &                   ARA \\
 2011 02 26       &- &     17.79 .10       &- &- &                    ARA\\  
 2011 07 30       &- &     18.08 .10        &- &- &               ARA  \\
 2011 08 02 &18.10 .07  & 18.07 .06 & 17.86 .04 &- &           Loiano \\
 2011 08 28       &- &     18.50 .09   &- &- &                 ARA  low quality \\
 2011 09 07 &18.23 .09 & 18.26 .06 & 18.07 .03 &- &     Loiano\\
 2011 09 21       &- &    18.35 .09  & 17.94 .12  &- &        ARA \\
 2011 10 02       &- &    18.22 .08    &- &- &                     ARA  \\ 
 2011 10 05       &- &    18.17 .10      &- &- &                     ARA  \\
 2011 10 29        &- &   18.22 .07     &- &- &                   ARA \\
 2011 11 15        &- &   18.36 .05 & 18.39 .04 & 18.17 .02  & Loiano \\
 2011 11 26        &- &   18.33 .08  &18.18 .06  &- &         ARA \\
 2011 12 17       &- &    18.45 .09    &- &- &                                 ARA  \\
 2011  12 23       &- &   18.51 .09    &- &- &                                 ARA \\
 2011 12 23        &- &   18.44 .06   &- &- &                                  ARA \\
 2011  12 26       &- &   18.42 .08 &  18.26 .08      &- &            ARA  \\
 2012 01 21        &- &   18.31 .09    &- &- &                                 ARA  \\
 2012 02 25        &- &   18.62 .12    &- &- &                                 ARA  \\
 2012 07 28       &- &   18.19 .09      &- &- &                           ARA \\   
 2012 08 24        &- &    18.42 .09    &- &- &                                 ARA \\
 2012 10 21 &18.39 .05  &18.49 .04 & 18.37 .05 &18.46 .08  & IAC80 \\
 2012 12 12 & 18.47 .06  &18.52 .06  & 18.30 .04   &- &             Loiano \\
 2012 12 30      &- &     18.29 .08    &- &- &                                ARA \\
 2013 01 07       &- &     18.48 .08   &- &- &                             ARA \\
 2013 02 04      &- &     18.38 .07    &- &- &                 ARA  \\
 2013 09 04       &- &      18.23 .10    &- &- &                                ARA  \\
 2013 10 01      &- &     18.68 .10      &- &- &                              IAC80 \\
 2013 10 02 & 18.62 .02  & 18.66 .03  & 18.46 .03 &18.81 .18 & IAC80 \\
 2013 10 04 & 18.62 .03  & 18.70 .03  & 18.46 .03 &18.71 .07&  IAC80\\
 2013 10 05 &18.60 .02  & 18.67 .02   &18.45 .03 &18.69 .07  & IAC80 \\
 2013 11 26 &18.60 .05 &  18.72 .06  & 18.43 .05   &- &               Loiano  \\
 2013 12 23      &- &      18.88 .14  & 18.40 .12 &- &                 ARA\\
 2013 12 23       &- &- &                   18.40 .12  &- &                ARA \\
 2013 12 30       &- &    18.68 .13   &- &- &                                  SSON\\
 2014 01 01       &- &-    &               18.49 .04   &- &                Loiano  \\
 2014 01 25 &18.65 .07  & 18.68 .04 & 18.60 .04     &- &               Loiano\\
 2014 08 08    &- &-     18.74 .10     &- &- &                      ARA \\
 2014 08 31    &- &-        18.79 .10   &- &- &                       ARA \\
 2014 09 27   &- &-         &              18.97 .13    &- &                ARA  \\
 2014 10 04   &- &         18.76 .05& 18.65 .03  &- &                  Loiano \\
 2014 10 20    &- &         18.87 .12     &- &- &                                ARA  \\
 2014 11 21    &- &        18.75 .12      &- &- &                               ARA  \\
 2014 12 11 & 18.67 .07 &  18.78 .07&  18.55 .05   &- &                Loiano \\
 2015 08 21    &- &        18.77 .08     &- &- &                               ARA  \\
 2015 11 06    &- &        18.78 .08     &- &- &                               ARA  \\
 2015 11 18    &18.70 0.03 &  18.84  .04     &18.50 0.04 &- &  Loiano$^1$  \\
 2015 12 02    &18.64 0.05 &  18.73  .05     &18.60 0.06 &18.74  0.07 &  Loiano  \\
  2015 12 10    &18.70 0.04 &  18.71  .04     &18.50 0.04 &- &  Loiano$^1$  \\
 \enddata
\tablenotetext{a}{ observatories: Greve: 30 cm telescope at Greve in Chianti (Firenze). ARA: 37 cm telescope of the Associazione Romana Astrofili at Frasso Sabino (Rieti). Loiano: 1.52 m telescope at the Loiano station of the Bologna Astronomical Observatory-INAF. SSON: the Sierra Stars Observatory Network in Mount Lemmon (Arizona). IAC80: 80 cm Tenerife telescope of the  Astrofisico de Canarias.   }
\tablecomments{ $^1$On November 14, 2015,  U=17.98 $\pm$0.08, ~on December 10, 2015,  U=17.70 $\pm$0.1}
\end{deluxetable}

\clearpage

\begin{deluxetable}{lcccccclll}
\tabletypesize{\footnotesize}
\tablewidth{0pt}
\tablecaption{Emission line measurements in the spectrum of GR~290. \label{tbl_spevol}}
\tablehead{
 \colhead{date}    & \colhead{$V$}  & \colhead{$H\alpha$}  & \colhead{$H\beta$}  & \colhead{band}  & 
\colhead{468.6}  & \colhead{587.6}  & \colhead{sp.typ.}  & \colhead{Obs}     & \colhead{remarks}     }
\startdata
  (1)     &    (2)     &       (3)    &     (4)     &      (5)    &      (6)    &     (7)   &        (8)       &   (9)    &     (10)\\
1992 10 06 &  16.4* &     47.7 1.0 &   10.0 1.0 &             &           & 1.25 0.1 &  late-B &   CalarA  &  468 np         \\   
2002 10 04 &  17.98 &    113.0 3.0 &   23.0 0.4 &  30.0 3.0   & 12.0 3.0  & 18.0 2.0 &  WN10h  &   SAO-M   &                 \\   
2003 02 02 &  17.70 &     97.6 3.0 &   27.3 0.5 &  15.5 1.0   &  6.6 1.0  & 22.3 1.0 & WN10.5h &  Loiano   &                  \\  
2004 02 14 &  17.56 &    100.  2.0 &   26.4 1.0 &  11.  1.0   &  1.3 0.5  & 22.2 2.0 &  WN11h  &   Ekar    &    468 v.un.     \\  
2004 11 12 &  17.18 &     97.3 2.0 &   24.0 1.0 &   7.0 2.0   &           & 21.0 1.0 &  WN11h  &   SAO-M   &   468 n.m.	  \\  
2004 12 07 &  17.18 &    113.  2.0 &   27.0 0.5 &   6.5 0.7   &  1.4 0.3  & 23.7 0.5 &  WN11h  &   Ekar    &     468 uncer.	   \\ 
2005 01 13 &  17.26 &    111.  3.0 &   27.5 1.0 &   6.4 1.5   &  1.1 0.5  & 21.5 0.5 &  WN11h  &   Loiano  &   468 v.un..      \\ 
2005 02 06 &  17.24 &    112.5 0.5 &   26.0 0.3 &   7.6 1.0   &  1.4 0.3  & 25.5 0.7 &  WN11h  &   SAO-S   &	     	\\ 
2005 08 31 &  17.5  &    120.0 5.0 &   26.7 0.5 &  18.5 1.5   &  7.0 0.8  & 26.0 1.5 &  WN10h  &   SAO-S   &	     	\\
2005 11 08 &  17.6  &    150.0 7.0 &   36.7 1.0 &  30.3 0.5   & 11.0 0.7  & 34.5 0.8 &  WN9h   &    SAO-S  &              \\   
2006 08 03 &  18.43 &              &   25.1 0.7 &  37.5 1.0   & 11.2 0.5  &          &  WN8h   &    SAO-S  & 	        \\   
2006 09 29 &  18.4  &              &   22.1 0.5 &  31.1 0.5   &  9.0 0.3  &          &  WN8h   &    WIYN   &        \\  
2006 11 21 &  18.50 &    144.   7.0 &   32.  2.0 &  47. 5.    &  19. 3.0  &  33. 3.0 &  WN8h   &    Loiano &     low  \\  
2006 12 14 &  18.55 &    129.   7.0 &   27.2 1.0 &  39.9 0.5  & 17.5 0.3  & 36.1 1.5 &  WN9h   &    Loiano &	   \\ 
2007 01 29 &  18.57 &    119.   4.0 &   27.3 1.0 &  36.5 3.0  & 13.1 2.0  & 26.9 0.5 &  WN8h   &    Loiano & 	   \\ 
2007 08 10 &  18.5  &              &    24.7 0.5 &  44.6 0.5  & 16.8 1.0  &          &  WN8h   &    SAO-S  & 	   \\ 
2007 10 06 &  18.6  &              &    24.3 0.3 &  46.0 1.0  & 16.6 0.5  &          &  WN8h   &    SAO-S  &	    \\
2008 01 08 &  18.62 &              &    23.7 0.3 &  43.0 2.0  & 14.4 1.0  &          &  WN8h   &    SAO-S  &      \\
2008 01 10 &  18.62 &    109.6  1.0 &            &            &           & 26.3 0.3 &  WN8h   &    SAO-S  &      \\
2008 01 23 &  18.65 &    128.   5.0 &   29.0 3.0 &  55.8 5.0  & 27.2 3.0  & 30.0 2.0 &  WN8h   &    Loiano &     465 uncer. \\
2008 02 07 &  18.65 &    132.   4. 0 &  30.6 1.0 &  44.0 3.0  & 21.3 1.0  & 27.5 1.0 &  WN8h   &    Loiano &  	\\
2008 09 08 &  18.44 &    115.   4. 0 &  28.1 0.5 &  33.4 1.5  & 12.8 0.3  & 30.1 1.0 &  WN9h   &    Loiano &     \\
2008 12 04 &  18.31 &    126.   2. 0 &  25.0 0.5 &  36.6 0.5  & 13.1 0.3  & 28.8 0.5 &  WN9h   &    WHT    &       \\
2009 02 15 &  18.36 &    117.   3. 0 &  27.9 0.3 &  38.5 2.0  & 12.8 0.5  & 31.   2. &  WN9h   &    Loiano &       low q    \\
2009 10 09 &  18.36 &    115.2  1. 5 &  26.0 0.5 &  38.8 1.0  & 18.4 1.5  & 27.8 0.5 &  WN9h   &    SAO-S  &          \\
2010 01 21 &  18.38 &    114.   2. 0 &  27.7 0.4 &  32.5 1.5  & 13.8 0.7  & 26.2 0.4 &  WN9h   &    Loiano &     low q.   	\\
2010 12 06 &  17.66 &                &  26.2 0.3 &  20.1 0.5  &  7.6 0.5  &          &  WN10h  &    SAO-S  &		      	\\
2010 12 18 &  17.93 &    106.   2. 0 &  28.2 0.5 &  21.2 2.5  &  8.   2.  & 32.5 2.0 &  WN10h  &   Loiano  &      468 uncer.   \\ 
2011 01 03 &  17.95 &    114.   2. 5 &  29.0 0.5 &  16.6 1.5   & 6.9 0.5  & 25.6 0.3 &  WN10h  &   Loiano  &     \\
2011 08 03 &  18.07 &    114.5  1. 0 &  29.2 0.5 &  23.4 1.5   & 8.0 0.7  & 30.5 0.5 &  WN10h  &   Loiano  &	         \\
2011 09 20 &  18.26 &    122.   5. 0 &  30.8 0.3 &  31.7 1.0   &13.0 0.5  & 34.6 0.5 &  WN9h   &    Loiano &              \\
2012 12 09 &  18.50 &    111.   2. 0 &  23.5 0.3 &  39.0 2.0   &16.3 0.5  & 30.0 1.0 &  WN8h   &    Loiano &      \\
2013 11 27 &  18.72 &    98.5   1. 0 &  23.1 0.3 &  44.7 0.5   &21.4 0.5  & 23.2 0.5 &  WN8h   &    Loiano & \\
2014 01 01 &  18.68 &    94.2   2. 0 &  22.4 0.2 &  38.7 2.0   &17.6 0.5  & 20.3 1.0 &  WN8h   &    Loiano &    \\
2014 08 03 &  18.74 &                &  20.0 0.3 &  45.3 1.0   &14.4 0.1  &          &  WN8h   &    SAO-S  &        \\
2014 12 11 &  18.78 &    92.9   0. 7 &  21.1 1.0 &  44.5 0.5   &20.2 0.5  & 18.7 0.3 &  WN8h   &    Loiano &     \\
2014 12 30 &  18.74 &    94.0   2. 0 &  21.9 0.5 &  43.9 1.5   &20.0 0.5  & 20.8 1.0 &  WN8h   &    Loiano &      \\
2015 01 26 &  18.7  &     83.   2.   &  23.5 1.5 &  50.   5.   &24.1 1.5  & 18.4 1.0 &  WN8h   &    Loiano &     low q. \\
2015 02 19 &  18.61 &                &  19.6 1.0 &  42.1 5.0   &16.3 1.0  &          &  WN8h   &    SAO-S  &    low q. \\  
2015 12 02 &  18.64  &    88.3   0.8.   &  20.5 1.3 &  47.0  3.   & 21.7 1.1  & 22.4 0.7 &  WN8h   &    Loiano &     \\
2015 12 10 &  18.70  &    89.5  1.0 &  20.5 0.5 &  35.0 0.7  & 20.0 0.7  & 19.5 0.8 &  WN8h   &    Loiano &     \\
\enddata						       
\tablecomments{ 
1) year, month, day. (2) V magnitude (B for the 1992 spectrum). (3) EW and error of H$\alpha$ excluding the nearby [\ion{N}{2}] doublet. (4) EW and error for H$\beta$. (5) EW of the 4600-4700~\AA~, emission blend excluding He~I 4713~\AA. (6) EW of the He~II 4686 line. (7) EW of He~I 5876. (8) spectral type. (9) Observatory: CalarA.: Calar Alto Obs. Ekar: Cima Ekar (Padova Observatory). Loiano: Loiano Station (Bologna Observatory). SAO-M/-S: 6 m Special Astronomical Observatory equipped with MPFS (M) or SCORPIO (S) cameras. WHT: 4m William Hershel Observatory. WIYN 3.5 m Observatory (Kitt Peak).  (10) remarks: 486: He~II 4686~\AA~ line; 465: the 4600-4700~\AA~\, emission blend; np: not present; n.m.: not measurable; (v.) uncer: (very) uncertain; low q.: low quality spectrum (noisy, bad skv). } 
\end{deluxetable}					       

\clearpage

\begin{table}\centering
\caption{Spectral data of GR~290 used to construct the models. $\Delta \lambda$ is spectral resolution. S/N is
signal to noise ratio.
}
\label{tbl_logfile}
\bigskip
\begin{tabular}{lllcc}
\hline
  Date  & Obs.    & Spectral                   & $\delta \lambda$ [\AA]&  S/N  \\
        &         &  range                     &  &        \\
Oct 2002& SAO-M   & 4250-6700                  &      6                & 16    \\
Feb 2003& Loiano  & 3650-8600                  &     11       &       \\
Jan 2005& Loiano  & 4150-6600                  &      3.5             &       \\ 
Sep 2006& WIYN$^1$   & 4000-5100                  & &       \\
Oct 2007& SAO-S   & 4000-5700; 5700-7500 &      5; 5             & 40; 20\\
Dec 2008& WHT     & 3000-10 000                &  0.8; 1.8   &    30; 20    \\
Oct 2009& SAO-S   & 3700-7800                  &      5                & 35    \\
Dec 2010& SAO-S   & 4100-5800                  &      5                & 60    \\
Aug 2014& SAO-S   & 4000-5650                  &      5                & 20    \\
\hline
\multicolumn{4}{l}{1 -- data taken from Clark et al. (2012).}
\end{tabular}
\end{table}
			      
\clearpage						      
			
\begin{table}
\caption{Derived properties of Romano's star. H/He indicates the hydrogen number fraction relative to helium,  $f$ is the filling factor. }
\label{tbl_parmodel}
\bigskip
\begin{tabular}{llllllllccc}
\hline
Date     & V    &   Sp. & $T_{eff}$ & $R_{2/3}$        & $L_*10^5$   & $\log L_*$      &   $ \dot{M}_{cl}10^{-5}$ & f  & $v_{\infty}$& H/He\\
         &[mag] &  type &   [kK]   &[$\rm R_{\odot}$]& [$L_{\odot}$]&[$\rm L_{\odot}$]&[$ \rm M_{\odot}\,yr^{-1}$]&    &[$\rm km/s$]&      \\ 
   &    &    &    &    &     &   &     &    &             &            \\      
 Oct 2002& 17.98  &  WN10h  & 28.0     &  39    & 8.0      &  5.90              &    2.4    & 0.15  & $250\pm100$  &1.7 \\ 
 Feb 2003& 17.70 &WN10.5h & 27.5       &  44    & 9.5   &  5.98            &    2.4    & 0.15  & $250 ~ 50$   &1.7 \\ 
 Jan 2005& 17.24& WN11h & 23.5$^{1}$    &  61    &10.05      &  6.02$^{1}$      &    4.0  & 0.15& $250 ~50$   &1.7 \\           
 Sep 2006& 18.4  & WN8h     & 31.0     &  28    & 6.7     &  5.83            &    1.5    & 0.15& $250 ~ 100$  &1.7 \\
 Oct 2007& 18.6 & WN8h     & 33.3      &  23.8   & 6.3    &  5.8            &    1.9      & 0.15& $370 ~ 50$   &1.7 \\ 
 Dec 2008& 18.31& WN8h    & 31.5       &  28.5    & 7.2    &  5.86           &    2.3      & 0.15& $370 ~ 50$   &1.7 \\ 
 Oct 2009& 18.36& WN9h   & 32.0        &  28.4    & 7.5    &  5.875          &    2.0    & 0.15& $300 ~ 100$  &1.7 \\
 Dec 2010& 17.95& WN10h &  26.7        &  42       &  8.0      &  5.9            &    2.6       & 0.15& $250 ~ 100$  &1.7 \\
 Aug 2014& 18.74&  WN8h  &  33.0       &  22.5     &5.3     &  5.72           &    1.7      & 0.15& $400 ~ 100$  &1.7 \\     
         &      &       &          &                 &              &       &   &    &    &    \\
 Errors  &      &       &$\pm1.0$  &     &$\pm0.5$      &         &   $\pm0.3$               &$\pm0.05$& & $\pm0.2$   \\
\hline
         &      &       &          &                 &              &                 &                          &    &             &     \\
Sep 2010$^{2}$ &17.8& WN10h &  26  &  41.5           &       &   5.85          &   2.18    &0.25&  265        &1.5  \\ 
             &      &       &      &                 &              &                 &                          &    &             &     \\
\hline
\end{tabular}
\tablecomments {$^{1}$  errors for these values are given in the text; $^{2}$ -- data taken from Clark (2012).}
\end{table}
\clearpage						      			
			
\clearpage
		
\clearpage				
		
\clearpage				
			
\clearpage

\clearpage
		
\clearpage
		
\clearpage

\clearpage

\clearpage

\clearpage
		
\acknowledgments
This research has made use of the plate archives of the following astronomical observatories: Asiago, Hamburg, Heidelberg, Yerkes, Pulkovo, SAO.
We thank P. Dalle Ave for having performed the scans of the Asiago plates. The plate digitization project of Hamburg Observatory (D. Groote) was funded by DFG grant GR969/4-1. 
We thank the Istituto Astrofisico de Canarias (IAC) for the use of the 80 cm telescope during the 2012 and 2013 campaigns. We are grateful to Franco Montagni for providing us with his photometric observations of GR~290 at Greve and to the ARA team who helped with the photometric monitoring of the star.
The observations at the 6-meter BTA telescope were carried out with the financial support of the Ministry of Education and Science of the Russian Federation (agreement No. 14.619.21.0004, project ID RFMEFI61914X0004).
 O. M. acknowledges the grant of Dynasty Foundation and the Russian Foundation for Basic Research (projects no. 14-02-31247,14-02-00291).
This research has made use of the SIMBAD database, operated at CDS, Strasbourg, France.





 \appendix 
  \section{Appendix}
 The Hamburg  plates have four different types of Agfa emulsions and no filters. All plates were digitized in transparency mode at the Hamburg Observatory with Epson Expression 10000 XL scannners, at 2400 dpi step, giving a scale of 0.74 arcsec/pixel, in the framework of a legacy project (P.I. D. Groote).
 Given the fast telescope focal ratio, the images are substantially affected by coma but the region around GR~290 is fairly usable~for aperture photometry. 
We transformed the plate transparency into pseudo-intensity with the formula I=(U-B)/(T-B), where T is the transparency of a given pixel, U the average transparency of the unexposed plate and B the transparency in the center of saturated stars \citep{nesci05}. Then we performed aperture photometry on GR~290 with IRAF/apphot, using as calibration the brightest reference stars of the sequence given in \citet{vio06} (namely from ku-a to ku-h), in the range $B$=14.8-19.08. We adopted the $B$ ~ magnitude scale, as it is more similar to that of the old plate emulsions.
Further stars of the sequence were not usable~due to crowding or because too faint. The adopted aperture radius was 4 pixels (3 arcsec), comparable~to the FWHM of the stars.  For each plate, a linear fit between instrumental and nominal magnitudes was derived and found quite satisfactory, with a slope near to the ideal case of 1.00; from these fits we derived the magnitude of GR~290
   together with the rms deviation of the comparison stars from the calibration and the slope of the linear fit. 
To estimate the photometric accuracy of our magnitudes we measured also a field star of comparable~magnitude and used its deviation from the fit as a photometric error, which resulted to be about 0.2 mag.  

The Heidelberg plates were  digitized with the Heidelberg Nexscan F4100,  with a resolution from 100 to 200 pixels per millimeter.
 The plates  found in the Pulkovo archive have been digitized with a Canon 5D MarkII camera.    
The Yerkes plates were scanned with an EPSON 10000XL at 2400 dpi.

All the POSS plates and the Asiago blue plates (103aO emulsion + GG13 filter) showed calibration curves fully compatible with a linear relation between instrumental and reference magnitudes; the Asiago V plates (103aD emulsion + GG14 filter) instead showed a marked curvature, requiring a parabolic fit. We therefore added a further star in the comparison sequence, fainter than GR~290, in order not to extrapolate its magnitude. The only suitable star in the \citet{vio07}  sequence is M31a-262 with V=17.61.





\end{document}